\documentclass[lettersize,journal]{IEEEtran}

\usepackage{url}
\usepackage{verbatim}

\usepackage{cite}
\usepackage{amsmath,amssymb,amsfonts}
\usepackage{mathrsfs}
\usepackage[linesnumbered, ruled]{algorithm2e}
\usepackage{amsthm}
\usepackage{multirow}
\usepackage{multicol}
\usepackage{tabularx}
\usepackage{algorithmic}
\usepackage{booktabs}
\usepackage{graphicx}
\usepackage{float}  
\usepackage{subfigure}  
\usepackage{bm}
\usepackage{booktabs}
\usepackage{textcomp}
\usepackage{xcolor}
\usepackage{xpatch}
\usepackage{bbm}
\usepackage{siunitx}
\definecolor{MycolorBlue}{HTML}{0080FF}

\begin{document}

\title{Statistical QoS Provision in Business-Centric Networks\\}

\author{Chang Wu, Yuang Chen,~\IEEEmembership{Student Member,~IEEE}, and Hancheng Lu,~\IEEEmembership{Senior Member,~IEEE}
\thanks{Chang Wu, Yuang Chen, and Hancheng Lu are with  CAS Key Laboratory of Wireless-Optical Communications, University of Science and Technology of China, Hefei 230027, China (e-mail: \{changwu, yuangchen21\}@mail.ustc.edu.cn; hclu@ustc.edu.cn). Hancheng Lu is also with Deep Space Exploration Laboratory, Hefei 230088, China.}

}



\maketitle
\begin{abstract}
More refined resource management and Quality of Service (QoS) provisioning is a critical goal of wireless communication technologies. In this paper, we propose a novel Business-Centric Network (BCN) aimed at enabling scalable QoS provisioning, based on a cross-layer framework that captures the relationship between application, transport parameters, and channels. We investigate both continuous flow and event-driven flow models, presenting key QoS metrics such as throughput, delay, and reliability. By jointly considering power and bandwidth allocation, transmission parameters, and AP network topology across layers, we optimize weighted resource efficiency with statistical QoS provisioning. To address the coupling among parameters, we propose a novel deep reinforcement learning (DRL) framework, which is Collaborative Optimization among Heterogeneous Actors with Experience Sharing (COHA-ES). Power and sub-channel (SC) Actors representing multiple APs are jointly optimized under the unified guidance of a common critic. Additionally, we introduce a novel multithreaded experience-sharing mechanism to accelerate training and enhance rewards. Extensive comparative experiments validate the effectiveness of our DRL framework in terms of convergence and efficiency. Moreover, comparative analyses demonstrate the comprehensive advantages of the BCN structure in enhancing both spectral and energy efficiency.
\end{abstract}


\newtheorem{def1}{\bf Definition}
\newtheorem{thm1}{\bf Theorem}
\newtheorem{lem1}{\bf Lemma}
\newtheorem{cor1}{\bf Corollary}
\newtheorem{proposition}{\bf \emph{Proposition}}

\section{Introduction}\label{Intro}
The development of mobile communication technology has reached a new milestone with the conceptualization of sixth-generation (6G) networks \cite{dang2020should,noauthor_report_nodate}. These networks promise to deliver rich service capabilities and the integration of artificial intelligence (AI), expanding the landscape of mobile communication scenarios \cite{lu20206g,letaief2021edge}. This progression is not only technological but also business-driven, as new business models emerge that require advanced communication capabilities. The interplay between technological advancements and business needs creates a cycle where each propels the other forward.

Despite the remarkable advancements in air interface technology and network architecture, significant challenges persist, particularly in the practical deployment of these innovations. A major challenge is the narrow focus of much existing research, which often prioritizes specific performance metrics like peak rate or spectral efficiency, without fully considering the broader context of system performance \cite{yang20226g,kasi2022d}. For example, cell-free networks show significant potential for service differentiation mitigation and spectrum efficiency improvement by uniform spectrum utilization and intensive access points (APs) deployment. However, severe interference and increased power consumption are issues that cannot be ignored \cite{ammar2021user}. The limitation becomes particularly pronounced as application scenarios become more complex and diverse, desiring a more nuanced approach to network design and resource management \cite{sun2020service}. The key challenges include: 1) Ensuring On-Demand Quality of Service (QoS). Different businesses have unique QoS requirements like latency, reliability, and rates. Ensuring that these diverse requirements are met on-demand is a critical challenge, particularly as network architectures evolve towards more user-centered designs. 2) Resource Efficiency. Efficient utilization of spectrum and power, is another major concern in the era of emphasizing green communication.


These challenges highlight the need for a comprehensive approach that goes beyond optimizing individual performance metrics. A holistic perspective is required that considers the inter-dependencies between various network layers and the trade-offs involved in resource allocation. The ultimate goal is to develop a network architecture and implementation scheme that not only meets the technical specifications but also aligns with the economic and practical realities of deployment, ensuring that the benefits of advanced technologies are fully realized in real-world applications.


Extensive research has focused on resource allocation in user-centric networks (UCN) to enhance system capacity or resource utilization \cite{qin2022user,ammar2021user}. Optimization efforts for service capabilities concentrate on cross-user spectral efficiency or overall capacity by adjusting AP-UE associations, transmission power, and channel allocation \cite{li2024uplink,wu2021proportional,zhang2020user,ammar2021distributed,ammar2021downlink}. Among these, optimizing weighted sum rates represents a particularly valuable variant due to its consideration of inter-user fairness \cite{li2024uplink,ammar2021distributed,ammar2021downlink}. The study in \cite{wu2021proportional} reformulates the optimization of the logarithm of time-averaged rates into a series of weighted sum rate maximization problems. The feasibility of power and rate requirements given the available bandwidth resources is initially assessed in \cite{zhang2020user}, advancing the research towards practical applications. Considering the demand for energy-efficient transmission, the minimization of system total power consumption or the maximization of energy efficiency is widely studied, with the limitation of computing resources and AP power \cite{shi2020power,zhang2023reconfigurable}.

However, the aforementioned studies predominantly focus on the achievable rate under resource constraints as a single performance metric, neglecting the investigation of transmission reliability and its diversity. The study in \cite{wang2023user} takes the average QoS satisfaction rate as the optimization target, but only throughput and UE association costs are included in the QoS considerations. Inspired by economics, user-centric resource allocation, comprehensively considering both QoS and the importance of services provided to UEs is explored in \cite{chen2022toward}. In research on the provisioning of QoS in communication systems, effective capacity has been widely used to statistically evaluate the service capability under varying quality of service conditions \cite{wu2003effective,Yuang2023When,chen2024statistical}. However, most of these works only analyze the performance boundaries of the system, failing to provide practical scheme design guidance due to rigid models.

Recent research has begun to explore the use of machine learning, and specifically deep reinforcement learning (DRL), to cope with the complexity of communication provisioning models. DRL offers a promising approach for optimizing network parameters across multiple layers and agents, owing to its capability to capture a wide range of influencing factors by a black-box methodology. Additionally, by circumventing the challenging solution processes typical of traditional methods, DRL-based methods can optimize the overall system performance in terms of utility functions \cite{kasi2022d,tian2021multiagent}, such as integrated resource efficiency \cite{zhang2022intelligent,li2021secure}, or a composite measure of service capacity and resource consumption \cite{yu2023asynchronous,tsukamoto2023user,yu2023user}. However, its application in multi-access systems, which require careful design of collaboration between APs, is still in its infancy. The challenge lies in designing algorithms and frameworks that can converge in a large action space in time, ensuring that the solution can provide the necessary QoS while optimizing resource utilization.


To cope with the increasingly diverse business requirements and resource objectives, we propose the architecture of Business-Centric Networking (BCN) by extending the UCN concept\cite{ammar2021user,yang20226g}. To our knowledge, this represents the first work that designs the structure of wireless systems and controls its workings around business characteristics. Compared with the UCN, which only allocates wireless resources based on the channel and specific system limit goals, BCN adapts air interface technologies and network architectures based on business requirements and resource objectives to achieve QoS performance guarantees of businesses. Air interface technology refers to the mechanism including resource allocation and transmission parameter decision, while network architecture refers to the AP application method that is used to create the expected wireless environment.

The coexistence of business objectives with varying emphases on throughput, latency, and reliability metrics, along with diversified resource efficiency goals, necessitates flexible resource orchestration to provision radio channels with specific characteristics. These distinct channel characteristics require integration with optimal transmission parameters to achieve desired QoS provisioning with minimal resource consumption. In other words, upper-layer applications with heterogeneous QoS requirements and differentiated emphases on physical-layer resources drive our investigation into transmission scheme adaptation and intelligent resource orchestration. The evolution from previous generations toward 6G reflects a broader trend toward increasingly flexible and adaptive network solutions, designed to accommodate both the complexity of modern communication requirements and the dynamics of operational environments.

Given the different emphasis on frequency and power resources, we studied the impact of channel and power allocation on the construction of transmission environments. With the consideration of the AP sleep mechanism, they jointly determine the underlying resource occupation. In the channel environment depicted by resource supply, the transmission error probability and retransmission of data packets are jointly considered, as they jointly affect the QoS supply of transmission in a coupled manner. Statistical QoS performance is evaluated due to the almost impossibility of deterministic reliability \cite{zhang2018heterogeneous,chen2023streaming,musavian2015effective}. The research problem is difficult to solve through traditional optimization methods due to the large number of coupled influencing factors in BCN. Therefore, a structure combining heterogeneous actors DRL with traditional optimization is designed to achieve the optimization goal of resource utilization. Our main contributions can be summarized as follows:

\begin{itemize}
    \item We propose a business-centric network architecture from the perspective of cross-layer optimization to achieve scalable QoS supply. A cross-layer QoS supply framework is proposed to achieve optimal resource efficiency while providing statistical QoS guarantees, with a joint consideration of power and bandwidth resource allocation at the physical layer, transmission parameters at the link layer, and AP networking mode at the network layer. The queuing delay, delay violation, transmission error, and retransmission of data are incorporated into this framework to characterize QoS indicators including throughput, delay, and reliability.
    \item To manage the complex interdependencies between resource allocation and QoS provisioning, a collaborative DRL framework is constructed. The heterogeneous power actor and sub-channel (SC) actors representing multiple APs are collaboratively optimized under the guidance of a common critic. The coupled transmission parameters decision is appropriately embedded into the DRL framework due to its impact on resource utilization. In addition, a novel multi-thread based experience sharing (ES) mechanism is proposed to improve the speed of offline training and system rewards.
    \item Extensive comparative experiments on DRL frameworks and system solutions are conducted to verify the effectiveness of the novel DRL frameworks in terms of convergence and efficiency. Furthermore, a cross-layer QoS supply scheme is implemented through a large number of ablation experiments to verify the integrated advantages of BCN architecture in spectrum efficiency and energy efficiency.
\end{itemize}

The remainder of the paper is organized as follows. In Sections \ref{SysM} and \ref{S3: Cro-Lay Opt}, we present the downlink communication model with AP cooperation and formulate the resource utility optimization problem. The transmission parameters decision is addressed in Section \ref{S4: Solu for TPD}. The setting and details of the DRL algorithm for SC scheduling and power allocation are described in Sections \ref{S5: Setting for DRL} and \ref{S6: Algo for DRL}, respectively. The performance of the algorithm and the contributions of each component are thoroughly validated in Section \ref{S7: Perfo Eval}. Finally, the conclusions of this study are discussed in Section \ref{S8: Conclu}.

\section{System Model and Problem Formulation}\label{SysM}
In this section, we first introduce the BCN scenario, channel models, and service models that distinguish business types. Subsequently, a resource allocation and transmission parameter decision problem is formulated, aiming to optimize the utilization of system bandwidth and energy resources under the triad QoS constraints imposed by the services. Some important notations are summarized in Tab. \ref{tabPara1}.

\begin{table}
\begin{center}
\caption{List of Main Parameters.}
\label{tabPara1}
\renewcommand\arraystretch{1.3}
\begin{tabular}{| l | l | l | l |}
\hline
Notations & Description\\
\hline
$\mathcal{M}/m$, $ \mathcal{N}/n$, $ \mathcal{K}/k$ & The set / index of AP, UE and SC\\
\hline
$\lambda_n$ & Service rate requirement for CF or EF\\ 
\hline
$D_n^{\text{th}}$, $\widetilde{D}_n^{\text{q}}$ & Total, queuing latency requirement\\ 
\hline
$D_n^{\text{s}}$ ,$D_n^{\text{q}}$ & Service, queue latency variable\\
\hline
$\varepsilon_{n}^{\text{s}}$, $\varepsilon_0$, $\varepsilon_n$ & BLEP, threshold and LVP requirement\\
\hline
$[t]$, $T^{\text{s}}$, $T^{\text{RTT}}$& slot $t$, Slot length and static delay \\
\hline
$A_n(t)$, $S_n (t)$, $\widetilde{S}_{n}[t]$& Arrival, service process and capacity\\
\hline
$\Gamma_{n}^{k}[t]$, $R_{n}^{k}[t]$ & SINR and achievable rate of UE $n$ on SC $k$\\
\hline
$a_{m,n}^k$      &$a_{m,n}^k = 1$ if AP $m$ serve UE $n$ on SC $k$\\
\hline
$p_m^k$, $p^0$ & Power of AP $m$ on SC $k$ and static power \\
\hline
$ X_n$, $\theta_{n}$   & Count of transmissions and latency exponent \\
\hline
 $L_p^{(a)}$, $L_p^{(p)}$& Average packet size (in bits) for EF and CF\\
\hline
 $\eta_{EEE}$,$\eta_{ESE}$ & Effective Energy / Spectral Efficiency\\
\hline
$F_n^{\text{ec}}$ & Source packet rate of UE $n$ can be supported\\
\hline
\end{tabular}
\end{center}
\end{table}

\subsection{Scenario}
\begin{figure*}
    \centering
    \includegraphics[width=15cm]{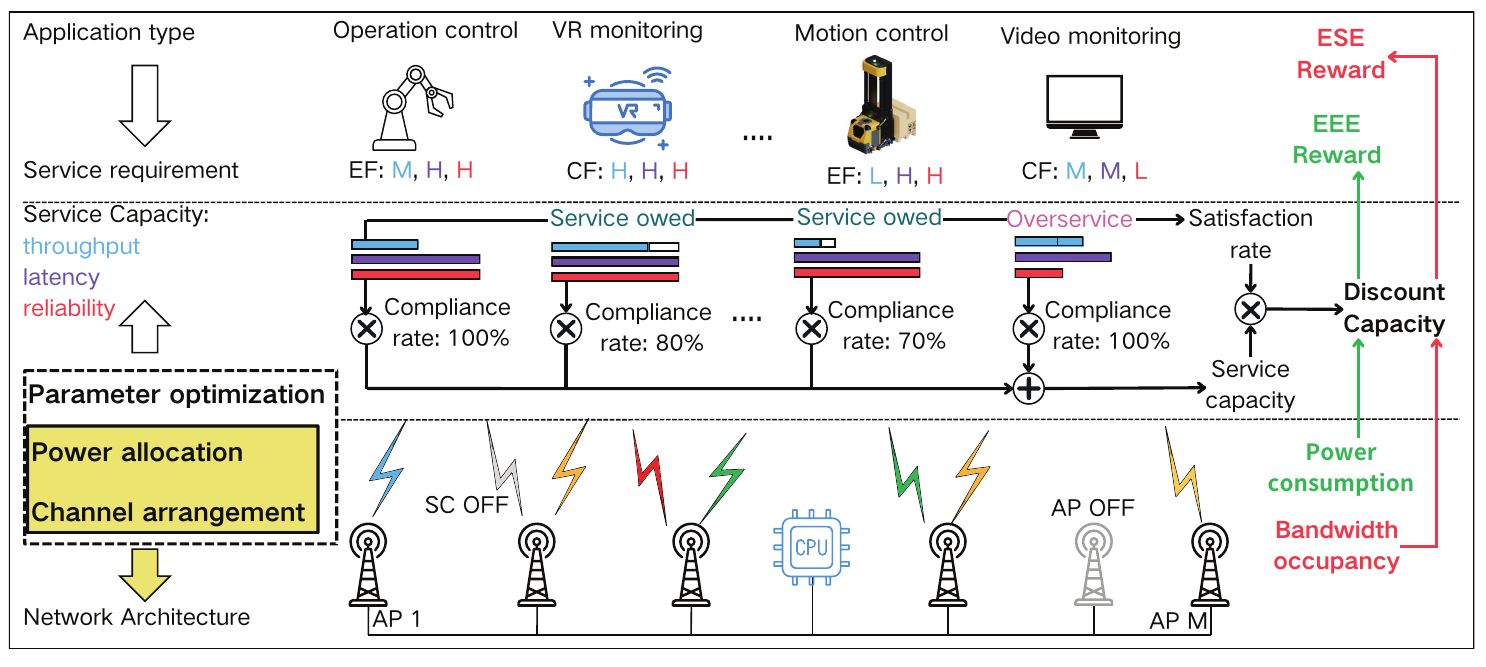}
    \caption{System Model. This diagram illustrates the on-demand services framework, where applications with diverse requirements are served by the underlying resources and facilities. The system rewards are calculated based on the completion of requirements and the consumption of resources.}
    \label{Fig:Scenario}
\end{figure*}

We start from a downlink UCN with one central processing unit (CPU), $M$ APs and $K$ User Equipments (UEs), as shown in Fig. \ref{Fig:Scenario}. Both APs and UEs are assumed to have a single antenna to focus on the core of our research.\footnote{Our analytical approach can be readily applied to the analysis of systems equipped with multi-antenna APs.} The sets of APs and UEs are denoted by $\mathcal{M}=\{1,2,\cdots,M\}$ and $\mathcal{N}=\{1,2,\cdots,N\}$, respectively. The total bandwidth $B$ is evenly partitioned into $K$ orthogonal SCs, denoted as $\mathcal{K}=\{1,2,\cdots,K\}$. Consequently, it holds that $B_0 = B/K$, where $B_0$ represents the bandwidth per SC. Each UE can be served by multiple APs on several SCs at the same time, which constitutes a flexible AP networking method. The transmission time is divided into slots of duration $T^{\text{s}}$.

We consider the deployment of the system within a communication landscape accommodating various applications with distinct QoS requirements. The system transmits either continuous periodic flows (CF) or event-driven aperiodic flows (EF) to devices represented as UEs, which conforms to the 3GPP evaluation criteria and holds significant practical implications\cite{3gpp_report_nr_ultra-reliable,3gpp_report_service}. In 6G-enabled factory settings, for instance, controllers transmit continuous video streams to devices while sporadically sending control information. We define the downlink packet rate for UE $n$ as $\lambda_n$, considering it successful if the packet is delivered to the UE within time $D_n^{\text{th}}$ with an error rate not exceeding $\varepsilon_n$. The triplet $\{\lambda_n, D_n^{\text{th}}, \varepsilon_n\}$ constitutes the entirety of the QoS requirements for UE $n$. Note that $\lambda_n$ represents a constant data rate for CF and mean value in EF, which will be further elucidated in the service model. Accordingly, all UEs are divided into CF business set $\mathcal{N}^{\text{p}}$ and EF set $\mathcal{N}^{\text{a}}$, with $\mathcal{N} = \mathcal{N}^{\text{p}} \cup \mathcal{N}^{\text{a}}$. 

\subsection{Physical Layer Model}



We define binary variable $a_{m,n}^k$ as SC allocation indicator, where $a_{m,n}^k = 1$ if $k$th SC is allocated to the $n$th UE from $m$th AP, and equals 0 otherwise. To mitigate severe interference, we postulate that each AP can only serve a single UE on each SC, which constitutes the sole restriction governing SC allocation and UE association within our system architecture. In essence, we permit the provision of service to a single UE by distinct APs across identical or distinct SCs. Furthermore, we allow different APs to cater to either the same or different UEs on different SCs. By introducing interference strategically and employing effective interference management, a substantial enhancement in resource utilization can be achieved, as elucidated in our results analysis.

The channel power gain from the $m$th AP to the $n$th UE on the $k$th SC, denoted by $h_{m,n}^k$, is formulated as $h_{m,n}^k=\alpha_{m,n} g_{m,n}^k$, where $\alpha_{m,n}$ and $g_{m,n}^k$ account for large-scale fading and small-scale fading, respectively. On the one hand, we assume that $\alpha_{m,n}$ can be promptly and accurately acquired or computed by the APs, remaining constant within one slot, and varying on a larger scale due to user movement. \footnote{The assumption is reasonable since the UEs' locations do not change too much in one slot, e.g. \cite{guo2019resource,wu2024cross}.} On the other hand, the ergodic Rayleigh fading in small-scale fading of each sub-channel is assumed in this paper, which implies that $g_{m,n}^k$ is an independent exponentially distributed random variable. With the power of AP $m$ on SC $k$ denoted as $p_m^k$, the signal-to-interference-to-noise ratio (SINR) of the signal received by UE $n$ on SC $k$ at slot $t$ can be calculated as
\begin{equation} 
    \Gamma_{n}^{k}[t] = \frac{\sum_{m \in \mathcal{M}} a_{m,n}^{k} p_m^k h_{m,n}^k[t]}{\sigma^2 + \sum_{m \in \mathcal{M}} \sum_{n'\in \mathcal{N}\backslash n} a_{m,n'}^{k} p_m^k h_{m,n}^k[t]},
\end{equation}
where $\sigma^2$ is the variance of Gaussian white noise at the receiver.

The majority of research endeavors characterize the transmission rate in terms of the Shannon limit rate denoted by a, $r=\log_2 (1+\gamma)$. This representation is applicable to general investigations concerning capacity limits but deviates from our specific focus, where we also place emphasis on the practical reliability of the transmission. Based on the finite blocklength coding theory \cite{sun2018optimizing,she2017cross,tang2019service}, we approximate the achievable rate (in \textit{bits/second}) of UE $n$ on SC $k$ as
\begin{equation}\label{rateInPhy}
    R_{n}^{k}[t] \approx \frac{B_0}{\ln{2}} \cdot \left[ \ln{\left(1+\Gamma_{n}^{k}[t]\right)} - \sqrt{\frac{V_n^k}{T^{\text{s}} B_0}}f_Q^{-1}(\varepsilon_n^{\text{s}}) \right],
\end{equation}
where $\varepsilon_n^{\text{s}}$ denotes the block error probability (BLEP) utilized to govern the reliability of the service process (i.e., single transmission process), affecting the achieved service rate due to limited resources and constrained channel quality. $f_Q^{-1}(a)$ is the inverse of the Gaussian Q-function
and $V_n^k$ denotes the channel dispersion expressed by $V_n^k = 1-\frac{1}{\left(1+\Gamma_n^k\right)^2} \approx 1$. The approximation is reasonable due to the generally excellent channel quality obtained by moderate path loss and appropriate interference management schemes in UCN \cite{sun2018optimizing}. Therefore, the achievable rate of UE $n$ at slot $t$ can be calculated as $R_n [t] = \sum_{k \in \mathcal{K}} R_n^k [t]$.
It is worth noting that $R_n [t]$ is highly volatile for given resource due to small-scale fading.

\subsection{Service Model}
\subsubsection{Event-driven Aperiodic Flow}
The random arrival of packets in EF, coupled with the fluctuation in service rates due to wireless channel conditions, drives the application of queuing theory in our study. Specifically, we assume that the packet arrival rate follows a Poisson distribution with parameter $\lambda_n$, a widely adopted assumption in the performance analysis of communication services \cite{she2017cross}. Additionally, the service time for packets is assumed to follow an exponential distribution due to the varying packet sizes and fluctuating service rates \cite{tian2021multiagent}, leading to the formation of an M/M/1 service model. Therefore, the probability of a aperiodic packet being dropped due to the queuing time $D_n^{\text{q}}$ exceeding the queuing delay threshold $\widetilde{D}_n^{\text{q}}$ can be calculated as
\begin{equation}\label{LVPofAperiodic}
    Pb\left( D_n^{\text{q}} > \widetilde{D}_n^{\text{q}} \right) = \exp\left\{- \left( \mathbb{E}_{\gamma}\left\{R_{n}[t] \right\} / L_p^{(a)} - \lambda_n \right) \widetilde{D}_n^{\text{q}} \right\},
\end{equation}
where $L_p^{(a)}$ is the average size (in \textit{bits}) of packet for aperiodic flow and $D_n^{\text{q}}$ is the random variable representing the steady-state latency experienced by data of UE $n$. The $ \mathbb{E}_{\gamma}\left\{\cdot \right\}$ denotes the statistical average of inner arguments with respect to service rate.

\subsubsection{Continuous Periodic Flow}
We analyze the evolution of a data transmission system based on the processes of data arrival and service. Consequently, given a static packet arrival rate $\lambda_n$ and slot length $T^{\text{s}}$, we define the arrival process 
$A_n[\tau] = \lambda_n \cdot \tau \cdot T^{\text{s}}$ for UE $n$ in the interval 
$[0,\tau T^{\text{s}}]$. And the amount of packets that UE $n$ can be served at the endpoint of time slot $t$ can be denoted as
\begin{equation}
	\widetilde{S}_{n}[t] = S_{n}\left[ t-1 \right] + T^{\text{s}}  R_n[t] / L_p^{(p)},
\end{equation}
 where $S_{n}[t]$ represents the cumulative service process of packets at the endpoint of slot $t$, and $L_p^{(p)}$ is the average size (in \textit{bits}) of packet for periodic flow. It is important to note that $S_{n}[t]$ is not equivalent to the amount of data that can be served to UE $n$, as transmission opportunities are wasted when the queue is empty. Therefore, the actual cumulative service process for UE $n$ is calculated as $S_{n}[t] = \min \left\{\widetilde{S}_{n}[t], A_{n}[t] \right\}$. We define the length of the packet queue for UE $n$ at the endpoint of time slot $t$ as $Q_n [t]$. The evolution of $Q_n [t]$ based on the aforementioned analysis is expressed as follows
\begin{equation}
    \begin{aligned}
        Q_n [t] 
        &= \left\{ Q_n [t-1] + \lambda_n T^{\text{s}} - R_n [t] T^{\text{s}} / L_p^{(p)}  \right\}^{+} \\
        &=A_n [t] - S_n [t] \geq 0 ,
    \end{aligned}
\end{equation}
where $x^+$ represents $\max (x,0)$, implying that the amount of data served will not exceed the amount of data arrived at any given moment.

With an allowed maximum queuing delay $\widetilde{D}_n^{\text{q}}$, the maximum queue length for UE $n$ can be computed as $\widetilde{Q}_{n} = \lambda_{n} \widetilde{D}_n^{\text{q}}$. Exceeding this queue length is considered a delay violation event, resulting in adverse effects on the user and thus representing one of the critical performance metrics requiring strict control. Consequently, the delay violation probability is defined as the ratio of the amount of data served within the delay violation time domain to the total data volume when the slot length is sufficiently large. Leveraging the SNC theory and the aforementioned block fading service characteristics, the effective capacity function of the UE $n \in \mathcal{N}^{\text{p}}$ can be calculated by
\begin{equation}
	EC_{n}^{\text{p}}\left( \theta_{n} \right) = -\frac{1}{\theta_{n}T^{\text{s}}} \ln{\mathbb{E}_{\gamma} \left\{ \exp \left[-\theta_{n}T^{\text{s}} R_{n}[t] / L_p^{(p)}\right]  \right\}}
\end{equation}
where $\theta > 0$ represents the delay exponent for UE $n$, governing the system's tolerance to delay violation events. In other words, based on the monotonically decreasing property of the effective capacity function with respect to $\theta$ under given spectrum and power allocations, $L_p^{(p)} \cdot EC_{n}^{\text{p}} (\theta_n = 0)=\mathbb{E}_\gamma \left[R_n\right]$ signifies the system tolerating unlimited delay violations, thereby achieving service capacity equal to channel capacity. Conversely, $EC_{n}^{\text{p}} (\theta_n = \infty)=0$ indicates the system's intolerance to any delay violation, resulting in zero service capacity. The effective capacity, guided by the parameter $\theta$, elucidates the source rate with statistical QoS guarantees under varying channel capacities. Consequently, when the source rate satisfies $\lambda_n = EC(\theta_n)$, the packet loss rate caused by delay violation for periodic flow can be calculated by
\begin{equation}
	\Pr\left\{ D_{n}^{\text{q}} > \widetilde{D}_{n}^{\text{q}} \right\} \approx \varphi_{n} \cdot \exp\left\{-\theta_{n} \lambda_{n} \widetilde{D}_{n}^{\text{q}}\right\},
\end{equation}
where $\varphi_{n} = \frac{\lambda_{n}L_p^{(p)}}{\mathbb{E}_{\gamma}\left\{R_{n}[t] \right\}} < 1$ represents the probability that the data queue of UE $n$ is non-empty in a stable state. It is noteworthy that data delivery exceeding the maximum queuing delay $\widetilde{D}_{n}^{\text{q}}$ are considered ineffective due to the immediacy of application data needs.

\section{Cross-Layer Optimization Scheme}\label{S3: Cro-Lay Opt}
In the context of the downlink scenario, the delivery latency of packets consists of queuing delay $D_n^{\text{q}}$ and service delay $D_n^{\text{s}}$, where the latter encompasses transmission delay $T^{\text{s}}$ and the retransmission delay involving several feedback-retransmission cycles. With a cumulative count of $X_n$ transmissions, the maximum latency for UE $n$ needs to satisfy
\begin{equation}
    \widetilde{D}_n^{\text{q}} + D_n^{\text{s}} = \widetilde{D}_n^{\text{q}} + T^{\text{s}} + \left( X_n -1 \right)\cdot  T^{\text{RTT}}  \leq D_n^{\text{th}},
\end{equation}
where $T^{\text{RTT}}$ denotes the static delay from transmission to receiving feedback, which is assumed to be two $T^{\text{s}}$ in this paper. The transmission error rate below the threshold $\varepsilon_0$ after a total of $X_n$ transmissions is considered to be completely eliminated by the forward error correction (FEC) mechanism \cite{kallel1994efficient}. Additionally, we assume that the resource consumption caused by retransmissions does not affect our analysis due to the tiny value of $\varepsilon_n^{\text{s}}$ and potential resource reservation

Extreme pursuit of system resource efficiency metrics might fail to satisfy service requirements. We therefore employ the concepts of Effective Energy Efficiency (EEE) and Effective Spectral Efficiency (ESE) to more comprehensively characterize wireless resource efficiency under guaranteed service quality. They are defined as the ratio of the effective capacity to the power and bandwidth, respectively, which can be formulated as follows
\begin{subequations}
    \begin{align}
        \eta_{EEE} &= \frac{\sum\limits_{n \in \mathcal{N}} \lambda_n}{\sum\limits_{m \in \mathcal{M}} \left[ \rho^{-1} \cdot \sum\limits_{k \in \mathcal{K}} p_{m}^{k}  + p^0 \cdot \mathbbm{1} \left\{ \mathbf{p}_m \right\}  \right]},\\
        \eta_{ESE} &= \frac{\sum\limits_{n \in \mathcal{N}} \lambda_n}{\sum\limits_{k \in \mathcal{K}} B^0 \cdot \mathbbm{1} \left\{ \mathbf{a}^k\right\}},
    \end{align}
\end{subequations}
where $\rho$ is the PA efficiency of the AP and $p^0$ denotes  the static power when the AP is power on. We define $\mathbbm{1} \left\{ \mathbf{a}\right\} = 1$ if matrix $\mathbf{a}$ is non-zero, and $\mathbbm{1} \left\{ \mathbf{a}\right\} =0$ otherwise, which signifies $\mathbbm{1} \left\{ \mathbf{p}_m \right\}$ and $\mathbbm{1} \left\{ \mathbf{a}^k \right\}$ as indicators characterizing the sleep status of AP $m$ and occupancy of SC $k$, respectively. The denominator of the EEE encompasses both static and transmission power components, which account for the dormancy state of the AP, affording a more accurate reflection of the ``power consumption per-bit'' concept in energy-efficient communications.

The EEE and ESE of the system are two interrelated performance metrics based on the above analysis, suggesting that we can explore the performance space of the system by studying the trade-off between EEE and ESE. Therefore, the normalization of EEE and ESE is expressed as
\begin{equation}\label{EEE and ESE}
    \eta_{EEE}^{\text{norm}} = \frac{\eta_{EEE}}{\eta_{EEE}^{\text{max}}}, \eta_{ESE}^{\text{norm}} = \frac{\eta_{ESE}}{\eta_{ESE}^{\text{max}}}.
\end{equation}
Subsequently, the system's utility function can be calculated as

\begin{equation}\label{total Unity}
    \mathcal{U} \left( a_{m,n}^k, p_m^k, X_n , \varepsilon_n^{\text{s}} \right) = \left(\eta_{EEE}^{\text{norm}}\right)^{\vartheta} \left(\eta_{ESE}^{\text{norm}}\right)^{1- \vartheta},
\end{equation}
where $\vartheta$ serves as a resource-preference factor quantifying the relative importance of EEE. This unity design adapts the Cobb-Douglas production function theory from microeconomics\cite{Douglas1976The}, whose multiplicative form has demonstrated effectiveness in characterizing synergistic multi-factor production, avoiding excessive deterioration of unilateral output. We aim to optimize the system utility by resource allocation at APs and transmission parameters for UEs. Then the problem can be formulated as follows
\begin{subequations}\label{Prob_Form0}
	\begin{align}
		&\max \quad  \mathcal{U} \left( a_{m,n}^k, p_m^k, X_n , \varepsilon_n^{\text{s}}\right) \\ 
		&\text { s.t. } \Pr\left\{ D_{n}^{\text{q}} > \widetilde{D}_{n}^{\text{q}} \right\} \leq \varepsilon_n,\quad  \forall n \in \mathcal{N} \label{Contraint: LVP}\\
        &\quad\quad EC_{n}^{\text{p}}\left( \theta_{n}, R_n \right) = \lambda_n,\quad  \forall n \in \mathcal{N}^{\text{p}} \label{Constraint: EC} \\ 
        &\quad\quad \widetilde{D}_{n}^{\text{q}} + D_{n}^{\text{s}} \leq D_{n}^{\text{th}}, \quad \forall n \in \mathcal{N} \label{Constraint: delay} \\
		&\quad\quad \left(\varepsilon_{n}^{\text{s}}\right)^{X_n} \leq \varepsilon_{0}, \quad \forall n \in \mathcal{N} \label{Constraint: TxErrRate} \\
		&\quad\quad \sum_{k\in \mathcal{K}}\!\! p_m^k \leq P_m, \quad \forall m \in \mathcal{M} \label{Constraint: Power}\\
		&\quad\quad a_{m,i}^k \cdot a_{m,j}^k =0, \forall i,j \in \mathcal{N}, i\neq j, \forall m \in \mathcal{M}, \forall k \in \mathcal{K} \label{Constraint:1AP1UEIn1SC}
	\end{align}
\end{subequations}
where \eqref{Constraint: EC}, \eqref{Contraint: LVP} and \eqref{Constraint: delay} denote the user's QoS requirements encompassing packet rate, reliability, and latency, respectively. The packet rate of EF is implicitly governed by delay violations, devoid of the additional packet rate constraints imposed by delay exponent in periodic traffic. Expression \eqref{Constraint: TxErrRate} is one of the important contents of our work, which achieves the QoS guarantee and maximizes resource efficiency based on the optimal combination of $\varepsilon_{n}^{\text{s}}$ and $X_n$. Expressions \eqref{Constraint: Power} encapsulates the constraints on the maximum transmit power for APs. Notably, \eqref{Constraint:1AP1UEIn1SC} serves as a pivotal constraint dictating that each AP remains singularly committed to servicing a sole UE on a designated SC.

From the resource allocation perspective, a higher degree of SC reuse could increase spectrum efficiency but may introduce greater inter-user interference, thereby lowering energy efficiency. From the perspective of transmission parameters decision and performance requirements, retransmissions can enhance the reliability of the transmission phase, thereby relaxing $\varepsilon_n^{\text{q}}$ in the queuing phase. However, the increased service (transmission and retransmission) delay reduces $\widetilde{D}_n^{\text{q}}$, posing higher performance requirements for the queuing. Larger $\varepsilon_n^{\text{s}}$ deteriorates transmission reliability while leading to higher transmission rates, which lead to more retransmissions due to the BLEP threshold $\varepsilon_0$. In summary, the EE and SE are mutually restricted and influenced by both transmission parameters and resource allocation.

The conclusion can be drawn that the problem is a mixed integer nonlinear programming problem and NP-Hard\cite{GUEROUT20171}. Analysis reveals that the coupling between resource variables and transmission parameters can be decomposed, resulting in two distinct sub-problems, namely the resource orchestration (RO) and the transmission parameter decision (TPD) sub-problem. The RO sub-problem consists of SC and power allocation. The $a_{m,n}^k$ can be optimized in the gradient direction of $\mathcal{U}$ by convex relaxation and gradient ascent. Based on fractional planning and the Lagrangian dual algorithm, the optimal $p_m^k$ under certain channels can be obtained. Alternating optimization of SC allocation and power allocation can approximate the global optimal solution, with computational complexity $\mathcal{O}(T_{\text{max}} \cdot (MNK+NX_{\text{max}}+MK))$ that is impractical in real-time work, where $T_{\text{max}}$ and $X_{\text{max}}$ is maximum number of iterations and transmissions, respectively. Therefore, we propose the HADRL scheme to solve the RO for its lower computational density in forward decisions.

\section{Solution for Transmission Parameter Decision}\label{S4: Solu for TPD}
The optimal QoS provision under the given resource arrangement rather than the ultimate resource efficiency goal is studied, as QoS may not be met due to a lack of resources or poor channel quality. We choose the supported source rate under the reliability and latency requirements as the objective of the sub-problem. The TPD sub-problem for periodic business can be represented as
\begin{subequations}\label{TPD1}
	\begin{align}
		\mathrm{\mathcal{P}1}:
		&\max_{\theta_{n}, \varepsilon_{n}^{\text{s}}, X_n}  -\frac{1}{\theta_{n}T^{\text{s}}} \ln{\mathbb{E}_{\gamma} \left\{ \exp \left[-\frac{\theta_{n}T^{\text{s}} R_{n}[t]}{L_p^{\text{p}}}\right]  \right\}}  \\ 
		&\text { s.t. } \frac{\lambda_{n} L_p^{\text{p}}}{\mathbb{E}_{\gamma}\left\{R_{n}[t] \right\}} \exp\left\{-\theta_{n} \lambda_{n} \widetilde{D}_{n}^{\text{q}}\right\} \leq \varepsilon_n, \forall n \in \mathcal{N}^{\text{p}} \label{TPD11Contraint: LVP}\\
		&\quad\quad \widetilde{D}_{n}^{\text{q}} + D_{n}^{\text{s}} \leq D_{n}^{\text{th}},\forall n \in \mathcal{N}^{\text{p}} \label{TPD11Constraint: delay} \\
		&\quad\quad \left(\varepsilon_{n}^{\text{s}}\right)^{X_n} \leq \varepsilon_{0}.\forall n \in \mathcal{N}^{\text{p}} \label{TPD11Constraint: TxErrRate}
	\end{align}
\end{subequations}
The problem can be explained as the optimization of effective service capabilities with the triplet of QoS requirements imposed on a given service system. Correspondingly, the TPD sub-problem of aperiodic traffic can be expressed as
\begin{subequations}\label{TPD2}
	\begin{align}
		\mathrm{\mathcal{P}2}:
		&\max_{\varepsilon_{n}^{\text{s}}, X_n} \quad \tilde{\lambda}_n  \\ 
		&\text { s.t. } \exp\left\{- \left( R_n - \tilde{\lambda}_n L_p^{\text{a}} \right) \widetilde{D}_n^{\text{q}} / L_p^{\text{a}} \right\} \leq \varepsilon_n, \forall n \in \mathcal{N}^{\text{a}} \label{TPD12Contraint: LVP}\\
		&\quad\quad \widetilde{D}_{n}^{\text{q}} + D_{n}^{\text{s}} \leq D_{n}^{\text{th}}, \forall n \in \mathcal{N}^{\text{a}} \label{TPD12Constraint: delay} \\
		&\quad\quad \left(\varepsilon_{n}^{\text{s}}\right)^{X_n} \leq \varepsilon_{0},\forall n \in \mathcal{N}^{\text{a}} \label{TPD12Constraint: TxErrRate}
	\end{align}
\end{subequations}
where $\tilde{\lambda}_n$ denotes the supportable source packet rate under given average service rate, which may exceed or fail to meet the requirement $\lambda_n$. Similarly, problem $\mathrm{\mathcal{P}2}$ characterizes the average source rate with QoS guarantee that a wireless system can support with fluctuating capacity given bandwidth and power resources.

The sub-problems $\mathrm{\mathcal{P}1}$ and $\mathrm{\mathcal{P}2}$ still pose MINLP problem with coupled variables $\varepsilon_{n}^{\text{s}}$ and $X_n$. However, the problem can be further decomposed due to the relatively small range of values for the transmission count. Once the transmission count is determined, the decoding error probability $\varepsilon_{n}^{\text{s}}$ and service latency $D_n^{\text{s}} = 2\left( X_n -1 \right) T^{\text{s}} $ become explicit, consequently influencing the availability of service rate $R_n$ and maximum queuing latency $\widetilde{D}_{n}^{\text{q}}$.

As concluded from our previous work \cite{wu2024cross}, relaxing both the maximum queuing latency $\widetilde{D}_{n}^{\text{q}}$ or service latency $D_n^{\text{s}}$ within the total budget $D_n^{\text{th}}$ tends to enhance service performance. Therefore, the optimal solutions for both sub-problems $\mathrm{\mathcal{P}1}$ and $\mathrm{\mathcal{P}2}$ must lie at the point where the constraint \eqref{TPD11Constraint: delay} and \eqref{TPD12Constraint: delay} are satisfied with equality. On the other hand, we infer from \cite[Theorem 1]{guo2019resource} that sub-problem $\mathrm{\mathcal{P}1}$ satisfies the constraint \eqref{TPD11Contraint: LVP} with equality if it is feasible, which also holds for constraint \eqref{TPD12Contraint: LVP} due to the monotonic increasing nature of $\varepsilon_n$ with respect to $\lambda_n$ in \eqref{LVPofAperiodic}. Consequently, the parameter $\theta_n$ in sub-problem $\mathrm{\mathcal{P}1}$ can be reformulated as
\begin{equation}\label{theta calculation}
	\begin{aligned}
		\theta_n
		&\overset{\eqref{TPD11Contraint: LVP}}{=} - \frac{1}{\lambda_{n} \widetilde{D}_{n}^{\text{q}}} \ln\left\{ \frac{\varepsilon_n \mathbb{E}_{\gamma}\left\{R_n [t]\right\}}{\lambda_n L_p^{\text{p}}} \right\} \\
		&\overset{\eqref{TPD11Constraint: delay}}{=}- \frac{1}{\lambda_{n} \left[ D_{n}^{\text{th}} - \left(2X_n - 1 \right) T^{\text{s}} \right]} \ln\left\{ \frac{\varepsilon_{n}\mathbb{E}_{\gamma}\left\{R_n [t]\right\}}{\lambda_{n} L_p^{\text{p}}} \right\}.
	\end{aligned}
\end{equation}
And the optimization objective in sub-problem $\mathrm{\mathcal{P}2}$ can be reformulated as
\begin{equation}
	\begin{aligned}
		\lambda_n 
		&\overset{\eqref{TPD12Contraint: LVP}}{=} \frac{\mathbb{E}_{\gamma}\left\{R_n [t]\right\}}{L_p^{\text{a}}} + \frac{\ln{\varepsilon_n}}{\widetilde{D}_{n}^{\text{q}}} \\
		&\overset{\eqref{TPD12Constraint: delay}}{=}  \frac{\mathbb{E}_{\gamma}\left\{R_n [t]\right\}}{L_p^{\text{a}}} + \frac{\ln{\varepsilon_n}}{D_{n}^{\text{th}} - \left(2X_n - 1 \right) T^{\text{s}}}.
	\end{aligned}
\end{equation}
Therefore, under the specified $X_n$, we transform sub-problems $\mathrm{\mathcal{P}1}$ and $\mathrm{\mathcal{P}2}$ into a unified formulation
\begin{subequations}\label{TPD3}
	\begin{align}
		\mathrm{\mathcal{P}3}:
		&\max \quad F_n^{\text{ec}} \left( \varepsilon_{n}^{\text{s}} \right) \\ 
		&\text { s.t. } \varepsilon_{n}^{\text{s}} \leq \left(\varepsilon_{0}\right) ^ {1 / X_n},\forall n \in \mathcal{N} \label{TPD2Constraint: TxErrRate}\\
		&F_n^{\text{ec}} = -\frac{1}{\theta_{n}T^{\text{s}}} \ln{\mathbb{E}_{\gamma} \left\{ \exp \left[-\frac{\theta_{n}T^{\text{s}} R_{n}[t]}{L_p^{\text{p}}}\right]  \right\}} , \forall n \in \mathcal{N}^{\text{p}} \label{TPD2ExpressionFecPeriodic}\\
		&F_n^{\text{ec}} = \frac{\mathbb{E}_{\gamma}\left\{R_n [t]\right\}}{L_p^{\text{a}}} + \frac{\ln{\varepsilon_n}}{ D_{n}^{\text{th}} - \left(2X_n - 1 \right) T^{\text{s}}}, \forall n \in \mathcal{N}^{\text{a}} \label{TPD2ExpressionFecAperiodic}
	\end{align}
\end{subequations}
where $F_n^{\text{ec}}$ represents the effective capacity $EC_{n}^{\text{p}}$ of periodic business, and the average source packet rate $\tilde{\lambda}_n$ that can be supported in aperiodic traffic. Following from above analysis, we further present a property of the optimal solution $\varepsilon_n^{\ast}$ to the problem $\mathrm{\mathcal{P}3}$ in what follows by Theorem 1.
\begin{thm1}\label{thm1}
	If the problem $\mathrm{\mathcal{P}3}$ is feasible, the optimal solution to the problem $\mathrm{\mathcal{P}3}$ satisfies $\varepsilon_n^{\text{s}\ast} = \left(\varepsilon_0\right)^{1/X_n}$.
\end{thm1}
\begin{proof}
	The preliminary conclusion is that $f_Q^{-1} \left(\varepsilon_n^{\text{s}}\right)$ decreases monotonically with $\varepsilon_n^{\text{s}}$ based on the properties of Gaussian Q-function. So the $R_n^k[t]$ in \eqref{rateInPhy} and therefore the $R_n [t]$, is monotone increasing functions of $\varepsilon_n^{\text{s}}$.

	For periodic traffic, $\theta_n$ demonstrates a monotonically decreasing trend concerning $R_n [t]$ leveraging expression \eqref{theta calculation}. Additionally, $F_n^{\text{ec}}$ emerges as a monotonically increasing function of $R_n [t]$ and a decreasing function of $\theta_n$ from expression \eqref{TPD2ExpressionFecPeriodic}. Based on above analysis, we assert that $F_n^{\text{ec}}$ represents a monotonically increasing function with respect to optimization variable $\varepsilon_n^{\text{s}}$, thus concluding the proof in periodic traffic.
	
	For aperiodic traffic, leveraging the monotonicity of $R_n [t]$ with respect to $\varepsilon_n^{\text{s}}$, we readily establish the monotonically increasing characteristic of $F_n^{\text{ec}}$ with respect to $\varepsilon_n^{\text{s}}$, which completes the proof for EF traffic.
\end{proof}

The optimal BLEP parameter $\varepsilon_n^{\ast}$ can be calculated easily based on \textbf{Theorem \ref{thm1}}. Therefore, the optimal transmission parameters $\{\varepsilon_n^{\ast}, X_n^{\ast}\}$ under given resource constraints, along with their corresponding supportable source packet rates $F_n^{\text{ec}\ast}$, can be computed through the process outlined in \textbf{Algorithm \ref{alg:P3Solu}}. These results serve as crucial foundations for resource arrangement optimization.

\begin{algorithm}[t]
	\caption{Solution Procedure for Problem $\mathrm{\mathcal{P}3}$}\label{alg:P3Solu}
	\KwIn{Maximum value of $X_n$, Statistical channel quality, QoS requirement triplet $\{\lambda_n, D_n^{\text{th}}, \varepsilon_n\}$
		
		\quad \quad \quad Decoding PER threshold $\varepsilon_{0}$, LVP $\varepsilon_{n}$}
	
	\KwOut{\emph{The optimal value $F_n^{\text{ec}\ast}$ and optimal parameter $\left\{\varepsilon_n^{\text{s}\ast}, X_n^{\ast}\right\}$ for UE $n \in \mathcal{N}$}}
	
	Set the optimal value $F_n^{\text{ec}\ast} = -\infty$, 
	
	\For{$X_n =1 \textrm{ to } X_n^{\text{max}}$}{
		$\varepsilon_{n}^{\text{s}} = \left(\varepsilon_{0}\right) ^ {1 / X_n}$
		
		Calculate $F_n^{\text{ec}}$ by \eqref{TPD2ExpressionFecAperiodic} for Aperiodic traffic;
		
		Get $\theta_n$ by \eqref{theta calculation};
		
		Calculate $F_n^{\text{ec}}$ by \eqref{TPD2ExpressionFecPeriodic} for Periodic traffic;
		
		\If{$F_n^{\text{ec}} > F_n^{\text{ec}\ast}$}{
			$F_n^{\text{ec}\ast} = F_n^{\text{ec}}$;
			$\varepsilon_n^{\text{s}\ast} = \varepsilon_n^{\text{s}}$;
			$X_n^{\ast} = X_n$}
	}
	\Return{the optimal value $F_n^{\text{ec}\ast}$ and optimal solution  $\left\{\varepsilon_n^{\text{s}\ast}, X_n^{\ast}\right\}$.}
\end{algorithm}

\section{Setting for Heterogeneous Actors Deep Reinforcement Learning}\label{S5: Setting for DRL}
The objective of the RL algorithm is to find a near-optimal policy $\pi^{\ast}\{a|s\}$ of Markov decision process (MDP) through agent-environment interaction and policy improvement \cite{arulkumaran2017deep}. The scheme is extended to a collaborative heterogeneous actors structure due to the presence of two types of decisions in each AP and a common ultimate goal.


\subsection{States of DRL}
The state variables should comprehensively describe the current condition of the system, encompassing all critical information that impacts resource allocation decisions. However, excessive state variables can decelerate the learning rate of the agent and potentially lead to instability in the training process. Therefore, the state variables for the agent used in SC orchestration and power allocation are described as follows.

\emph{1) SC Actor State $s_c^t$: }We delineate state of SC actors at time step $t$ into three components: the SINR of each UE on each SC $\Gamma_{n}^{k}[t]$, the large-scale fading coefficient between each UE and each AP $\alpha_{m,n}$, and the QoS requirements triplet of each UE $\{\lambda_n, D_n^{\text{th}}, \varepsilon_n\}$. Thus, we have 
\begin{equation}
    s_c^t = \{\Gamma_{n}^{k}[t], \alpha_{m,n}[t], \lambda_n, D_n^{\text{th}}, \varepsilon_n, \forall m \in \mathcal{M}, \forall n \in \mathcal{N}\}.
\end{equation}

\emph{2) Power Actor State $s_p^t$: }We integrate SC decision actions into the power actor states since SC orchestration serves as a prerequisite for power allocation. Consequently, the state of the power actor is represented as
\begin{equation}
    s_p^t = \{s_c^t, a_c^t\}.
\end{equation}

\subsection{Action of DRL}
Upon acquiring the state of the environment, the agent's actions can be derived through the forward propagation process of the actor network. The definition of actions must encompass all possible behaviors of the agent within the environment, which constitutes the ultimate objective of our training.

\begin{figure}
    \centering
    \includegraphics[width=8cm]{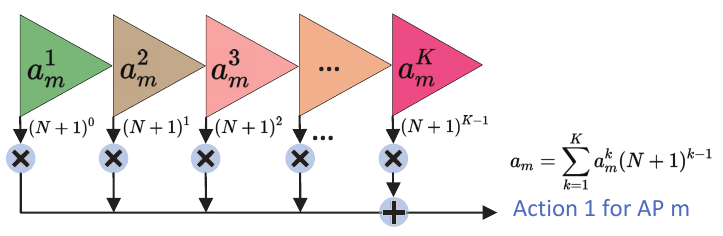}
    \caption{$K$-bit-$(N+1)$ number encoding for AP $m$}
    \label{Fig:action_1_decode}
\end{figure}

\emph{1) SC Actor Action $a_c^t$: }Since each AP serves either one UE or zero (off) per SC, we introduce a service indicator variable $a_m^k$ with its definition as
\begin{equation}
    a_m^k = 
    \begin{cases} 
    0, &  a_{m,n}^k =0, \forall n \in \mathcal{N}\\
    n, &  a_{m,n}^k =1
    \end{cases}
\end{equation}
where $a_m^k=0$ indicates the SC $n$ on AP $m$ is off. Each AP takes actions on $K$ SCs, where $(N+1)$ actions can be implemented on each SC, including $N$ UEs and being off. Therefore, we utilize a $K$-bit-$(N+1)$ number encoding to represent the action value for each AP, as illustrated in Fig. \ref{Fig:action_1_decode}. So we have
\begin{equation}
    a_c^t = \{a_m [t], \forall m \in \mathcal{M}\}
\end{equation}

\emph{2) Power Actor Action $a_p^t$: }Compared to discretizing power values into power levels, continuous power actions undoubtedly lead to finer action adjustments and closer-to-optimal performance. Therefore, we employ a continuous actor network to output the power ratios for all APs across all SCs. The output of the actor network for each AP will be conditionally processed through a soft-max layer to ensure that the sum of power ratios for each AP does not exceeds 1, which conforms to the power constraints of each AP. Therefore, the action of the power actor can be expressed as
\begin{equation}
    a_p^t = \{p_m^k, \forall m \in \mathcal{M}, \forall k \in \mathcal{K}\}.
\end{equation}

\subsection{Reward of DRL}
As a crucial basis for policy improvement, the reward design should accurately reflect the impact of actions on the ultimate objective in specific environmental states. Moreover, the reward density in the action space should be appropriately designed to avoid non-uniform variations and sparse rewards.

\emph{1) ESE Reward $r_s^t$: }
We introduce three elements to participate in the calculation of the ESE reward: the capacity satisfaction ratio for each user, the overall User demand satisfaction ratio in the system, and the system's SC utilization. Thus, the calculation process for the system's ESE reward is described as follows: (1) The penalized effective capacity for each user is derived from the effective capacity obtained in problem $\mathrm{\mathcal{P}3}$ and the completion rate $\omega_n^{\text{u}}$ (not exceeding 100\%): $\omega_n^{\text{u}} \times F_n^{\text{ec}}$. The sum of the values among all users yields the system's service capacity. (2) The discounted capacity for system is obtained from the system's service capacity and the user satisfaction rate $\omega^s$ within the system: $\omega^s \times \sum_{n \in \mathcal{N}} \omega_n^{\text{u}} F_n^{\text{ec}}$. (3) The system's ESE reward is calculated based on the discounted capacity and bandwidth consumption, as follows:

\begin{equation}
    r_s^t = \frac{\omega^s \!\!\sum\limits_{n \in \mathcal{N}} \omega_n^{\text{u}} F_n^{\text{ec}}}{\sum\limits_{k \in \mathcal{K}} B^0 \cdot \mathbbm{1} \left\{ \mathbf{a}^k\right\}},
\end{equation}
where $\mathbf{a}^k$ is the SC allocation vector performed by all APs on SC $k$. An additional interpretation for the $\omega_n^{\text{u}}$ and $\omega^s$ is that the $\omega_n^{\text{u}}$ is the proportion of available capacity for user $n$ to its service requirement, while $\omega^s$ is the proportion of users in the system whose available capacity meets service requirements.

\emph{2) EEE Reward $r_e^t$: }The only difference from the calculation of ESE rewards is that the penalty in the EEE rewards is realized by the power consumed. Thus, we have
\begin{equation}
    r_e^t = \frac{\omega^s \!\!\sum\limits_{n \in \mathcal{N}} \omega_n^{\text{u}} F_n^{\text{ec}}}{\sum\limits_{m \in \mathcal{M}} \left[ \rho^{-1} \cdot \sum\limits_{k \in \mathcal{K}} p_m^k  + p^0 \cdot \mathbbm{1} \left\{ \mathbf{p}_m \right\}  \right]}.
\end{equation}
Then, the final reward $r^t$ is calculated based on the same process as Eq. \eqref{EEE and ESE} and \eqref{total Unity}. Notably, each component of the reward is influenced by the actions of both actors and the transmission parameters, rather than being solely affected by a single one. This characteristic not only ensures that the system's reward encompasses all decision-making aspects but also presents significant challenges in the design of the algorithm's structure.

\section{Algorithm Structure for HADRL}\label{S6: Algo for DRL}
In this section, we present our proposed novel algorithm designed to expedite and stabilize the training process of HADRL through experience sharing. Taking into account the heterogeneous action space, which includes both discrete and continuous actions, we developed a heterogeneous actor architecture that executes synchronously, with PPO \cite{schulman2017proximal} serving as the backbone. 

\subsection{PPO Preliminary}
Proximal Policy Optimization (PPO) is a widely used reinforcement learning algorithm known for its balance between simplicity and performance. PPO falls under the category of policy gradient methods and is designed to optimize policies in a stable and efficient manner. The key idea behind PPO is to restrict the policy update step to ensure that the new policy does not deviate significantly from the old policy, thereby improving training stability.

PPO introduces a clipped surrogate objective function to constrain the policy update. The objective function for PPO can be defined as
\begin{equation}
    L^{CLIP}(\theta)=\mathbb{E}_{\pi_{\theta_{old}}}\left[ f\left(r_p^t(\theta), A^t\right)\right],
\end{equation}
where 
\begin{equation}\label{f_r_p_A_t}
    f\left(r_p^t(\theta), A^t\right) = \min \left(r_p^t(\theta) A^t, \operatorname{clip}\left(r_p^t(\theta), 1-\epsilon, 1+\epsilon\right) A^t\right).
\end{equation}
In \eqref{f_r_p_A_t}, $r_p^t(\theta) = \frac{\pi_\theta (a^t|s^t)}{\pi_{\theta_{old}} (a^t|s^t)} $ is the probability ratio between the new policy $\pi_\theta$ and the old policy $\pi_{\theta_{old}}$, $A^t$ is the advantage function. The clip function with hyperparameter $\epsilon$ restricts the ratio into the interval $[1-\epsilon, 1+\epsilon]$. The clipping function ensures that the policy update does not exceed a predefined threshold, thus preventing large policy updates that could destabilize training. The advantage function $A^t$ is a crucial component in PPO, providing an estimate of the relative value of taking a particular action in a given state.

PPO employs separate networks for the policy and value function, which are updated iteratively. The policy network is updated by maximizing the clipped surrogate objective $L^{CLIP}(\phi)$, while the value network is updated by minimizing the mean squared error between the predicted value and the observed returns. The robustness and sample efficiency of PPO make it a suitable choice for complex environments, including the resource allocation problem in heterogeneous multi-agent systems. 

\begin{figure}
    \centering
    \includegraphics[width=9cm]{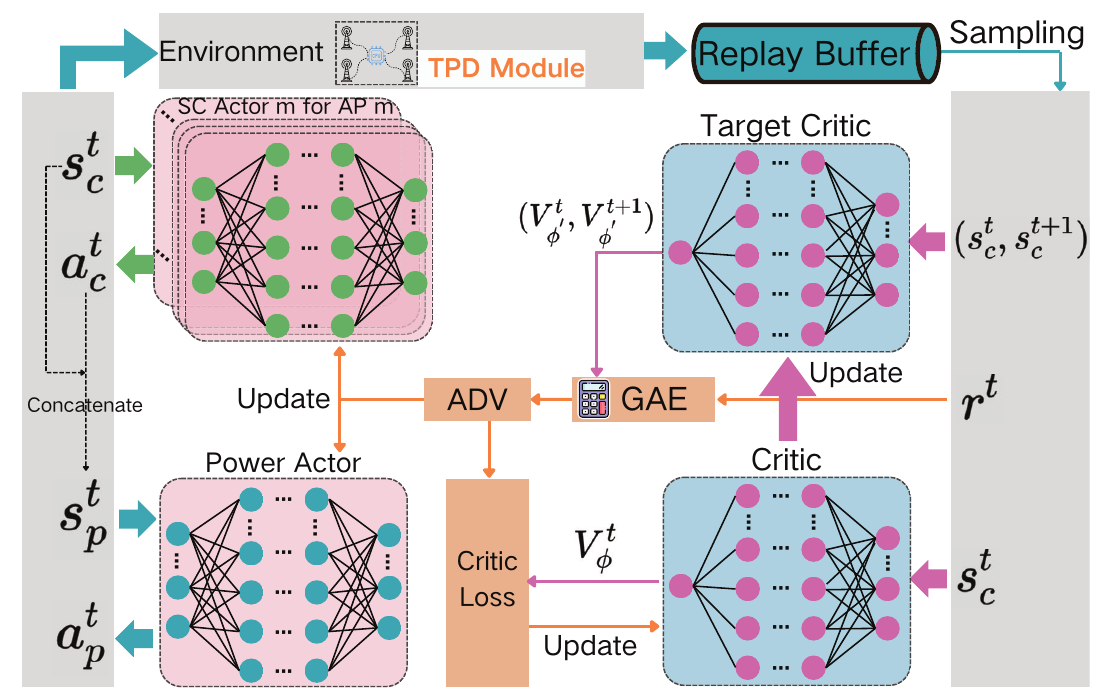}
    \caption{Collaborative Optimization with Heterogeneous Actors (COHA) structure.}
    \label{Fig: COHA}
\end{figure}

\subsection{Collaborative Optimization with Heterogeneous Actors (COHA)}
The common objective necessitates that the policy evolution of the two actors within the algorithmic framework is collaborative rather than competitive. Considering the significantly different roles and action spaces, a heterogeneous actor architecture has been designed in the proposed scheme. The two actors are updated separately based on a shared reward and distinct states, as shown in Fig. \ref{Fig: COHA}.

In particular, the initial state $s_c^t$ is sent into the $M$ SC actor networks at the beginning of each episode, generating $M$ SC allocation actions $a_c^t = ({a_1, \cdots, a_M})$. The actions for power $a_p^t$ are then produced by the power actor network, based on the integrated $s_c^t$ and $a_c^t$. The environment executes both actions along with Algorithm \ref{alg:P3Solu} to obtain the optimal transmission parameters, subsequently distributing the system reward $r^t$. The tuples $\{s_c^t, a_c^t, a_p^t, r^t, s_c^{t+1}\}$ are stored in the replay buffer and batch-sampled during network updates phase. During each update cycle, the SC states $s_c^t$ and the next state $s_c^{t+1}$ are sent to the target critic network to generate the state value function. They are passed to the GAE together with the rewards for advantage calculation. With the target critic network parameterized by $\phi^{'}$, the truncated Generalized Advantage Estimation (GAE) method is often used to compute $A^t$ due to its ability to balance bias and variance. which is defined as
\begin{equation}
    A^t = \sum_{l=0}^{T^{\text{b}}-1} (\gamma\lambda)^l \delta^{t+l},
\end{equation}
where $T^{\text{b}}$ denotes the length of the batch sample, $\delta^{t+l} = r^{t+l} + \gamma V_{\phi^{'}}(s^{t+l+1}) - V_{\phi^{'}}(s^{t+l})$ is the temporal difference error, $\gamma$ is the discount factor, and $\lambda$ is a factor that determines the bias-variance trade-off.

The objective function of PPO with entropy regularization is given by
\begin{equation}\label{L_PPO}
	    L^{PPO}(\theta)=L^{C L I P}(\theta)+c \cdot \mathbb{E}_{\pi_{\theta_{old}}}\left[H\left(\pi_\theta\left(\cdot \mid s^t\right)\right)\right]
\end{equation}
where $L^{C L I P}(\theta)$ denotes clipped surrogate objective function, $c$ is a coefficient that determines the weight of the entropy term, and $H(\cdot)$ denotes the entropy of the policy distribution. Combining the advantage function $A^t$, the policy gradient for $m$th SC actor and power actor is estimated by
\begin{subequations}
	\begin{align}
         \nabla \theta_c^m\! &=\! \frac{1}{T^{\text{m}}} \!\sum_{l=0}^{T^{\text{m}} - 1}\! \nabla_{\theta_c^m}\! \left[ f_l\!\left(r_p^t(\theta_c^m), A_l^t\right) +c \cdot H_l\left(\pi_{\theta_c^m}\right)\right] \label{Delta_cm}\\
         \nabla \theta_p &= \frac{1}{T^{\text{m}}} \sum_{l=0}^{T^{\text{m}} - 1} \nabla_{\theta_p} \left[ f_l\left(r_p^t(\theta_p), A_l^t\right) +c \cdot H_l\left(\pi_{\theta_p}\right)\right] \label{Delta_p}
	\end{align}
\end{subequations}
where $T^{\text{m}}$ is the length of mini-batch sample. Therefore, the actor networks can be updated via mini-batch stochastic gradient descent. Similarly, the gradient of critic network is calculated to update the critic network
\begin{equation}\label{L_VF}
    L^{VF}(\phi)=\mathbb{E}_{\pi_{\theta_{old}}}\left[\left(V_\phi(s^t) - V_{target}^t\right)^2\right],
\end{equation}
where $V_{target}^t = A^t + V_{\phi^{'}}(s^t)$. $V_\phi(s^t)$ and $V_{\phi^{'}}(s^t)$ are the value functions output by the critic and target critic respectively, with same structure but different parameter $\phi$ and $\phi^{'}$.
The parameters of the target network are covered by the critic network after every $C$ update to enhance the stability of the update process.The overall training process of COHA-based SC orchestration and power allocation scheme is shown in Algorithm \ref{alg:COHA}.

\begin{algorithm}[t]
	\caption{COHA-based DRL Framework for Resource Orchestration}\label{alg:COHA}
	\textbf{Initialize} the system environment $s_c^0$, SC and power actors network parameter $\theta_c^m, \forall m \in \mathcal{M}, \theta_p$, critic and target network parameters $\phi, \phi^{'}$ and training parameters: learning rate $l_c^m, l_p$;
	
	\For{$\textit{time step } t=1, 2, \cdots$}{
 
        \For{SC actor $m = \{1,2,\cdots, M\}$}{
            Input $s_c^t$ to obtain SC action $a_m[t]$ according to $\pi_{\theta_c^m}$;
        }
        Concatenate $\{s_c^t, a_c^t\}$ and input to power actor, obtain power action $a_p^t$ according to $\pi_{\theta_p}$;
        
        Execute action $\{a_c^t, a_p^t\}$ in environment, execute \textbf{Algorithm\! \ref{alg:P3Solu}}, obtain reward $r^t$ and next state\! $s_c^{t+1}$;

        Save the tuples $\{s_c^t, a_c^t, a_p^t, r^t, s_c^{t+1}\}$ in buffer, and $s_c^t \leftarrow s_c^{t+1}$;

        \If{$t\%T^{\text{b}} == 0$}{
            Sample $\{s_c^t, a_c^t, a_p^t, r^t, s_c^{t+1}\}$ till end, compute Advantage $A^t$;

            Compute SC actor gradient \! $\nabla_{\theta_c^m}$\! by Eq. \!\eqref{Delta_cm}\! and update

            $\theta_c^m \leftarrow \theta_c^m - l_c^m \nabla_{\theta_c^m}L^{PPO}(\theta_c^m), \forall m \in \mathcal{M}$;

            Compute power actor gradient \! $\nabla_{\theta_p}$\! by Eq. \!\eqref{Delta_p}\! and update

            $\theta_p \leftarrow \theta_p - l_p \nabla_{\theta_p}L^{PPO}(\theta_p)$;

             Update critic with MSE loss using Eq. \eqref{L_VF};

             Clear the replay buffer;
        }

        \If{$t\%\textit{ episode length} == 0$}{
        Reset environment and get new state $s_c^t$;
        }
        
        Assign target network $\phi^{'} \leftarrow \phi$ every $C$ steps;
	}
\end{algorithm}

\subsection{Experience Sharing based on Multi-threading}
\begin{figure}
    \centering
    \includegraphics[width=8cm]{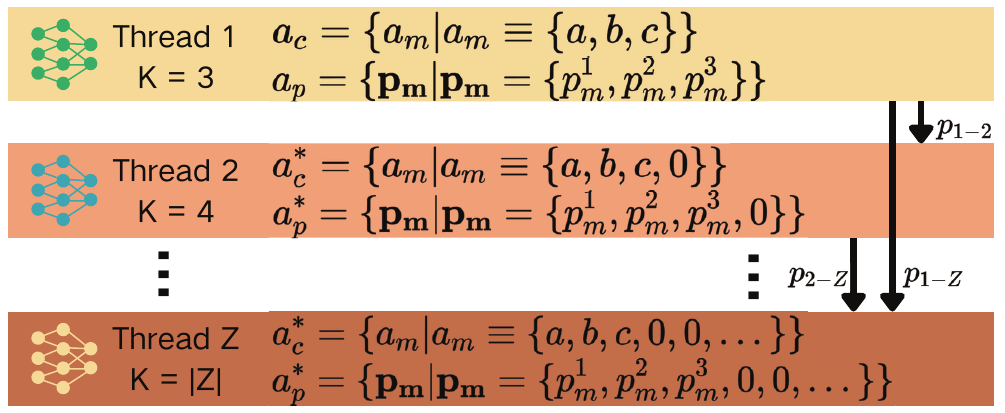}
    \caption{Multi-thread based experience sharing. The arrow with probability $p_{x-y}$ represents the sampling probability of high reward experience from thread $x$ to thread $y$, and $a^{\ast}$ represents the action suitable for this thread that evolved from action $a$. The symbol `'$\equiv$'' indicates equivalence.}
    \label{Fig:multi-thread}
\end{figure}

The output dimension of the $m$th SC actors, denoted as $\lvert a_m \rvert = (N+1)^K$, increases dramatically with the number of SCs. Exploring actions that result in minimal system bandwidth usage and therefore maximizing bandwidth rewards becomes a tricky business, as their proportion within the action space diminishes. Inspired by heterogeneous multi-threading algorithms \cite{mnih2016asynchronous}, we address this issue by asynchronously executing multiple threads with varying SC counts and sharing training experiences among them. Specifically, when multiple threads with different SC counts train concurrently, those with fewer SCs tend to discover high-reward action combinations more readily and converge faster. We probabilistically sample high-reward action combinations $\{a_c,a_p\}$ from threads with smaller SC counts and inject them into threads with larger SC counts by zeroing out the power on the extra SCs, as illustrated in Fig. \ref{Fig:multi-thread}. Consequently, the training processes with larger SC counts are influenced to converge more rapidly to higher system rewards.

\subsection{Convergence Analysis}
The theoretical guarantee of the convergence of the scheme is first introduced. Since PPO is used as the benchmark algorithm, its convergence is based on the following theory. PPO can converge to a local optimum when the following conditions are satisfied: (1) the policy is parameterized appropriately; (2) the learning rate sequence satisfies the Robbins-Monro condition; (3) the reward function is bounded. In addition, the Clip mechanism ensures the smoothness of policy updates and avoids divergence. All actors share the value evaluation of the same Critic in our proposed COHA-ES framework, establishing a unified optimization direction. Policy thrashing is avoided by implementing $\mathbb{E}[D_{KL}(\pi_{\text{old}} || \pi_{\text{new}})] \leq \delta$ with a multi-threaded experience pool. In addition, as part of the state transition function f (s, a), Algorithm 1 does not affect the policy gradient calculation. As part of the environmental feedback, the output of Algorithm 1 is only required to be observable, without breaking the Markov property.

\subsection{Complexity Analysis}
Focusing on real-time deployment requirements, we first analyze the complexity of the inference stage with $M$ SC Actors and one Power Actor. Each fully-connected layer in the Actor networks has complexity $\mathcal{O}(W_{in} \times W_{out})$, yielding total $L$-layer complexity $\mathcal{O}(L \times W^2)$. The collective computation for $M$ SC Actors is $\mathcal{O}(M \times L \times W^2)$. Taking into account the computational cost $\mathcal{O}(L \times W^2)$ of the Power Actor and the complexity $\mathcal{O}(N)$ of \textbf{Algorithm \ref{alg:P3Solu}}, which is connected to DRL after each decision, the total inference complexity is $\mathcal{O}((M+1)L W^2 + N)$. Parallelization can be implemented in practical deployments to reduce overall complexity to $\mathcal{O}(L W^2 + N)$. The complexity increases linearly only with the number of UEs, which demonstrates the advantages of the DRL scheme integrated with the optimization algorithm in terms of computational density and scalability.

The complexity of backward propagation in the Critic network is $\mathcal{O}(LW^2)$. Actor network updates occupy $\mathcal{O}((M+1)LW^2)$ complexity and experience replay incurs $\mathcal{O}(B_a \times LW^2)$ complexity, where $B_a$ denotes the batch size. Our framework employs multi-threaded experience sharing, introducing communication overhead $\mathcal{O}(M \times |\theta|)$, with $|\theta|$ as parameter dimensionality. Consequently, the total training complexity is $\mathcal{O}(T_{\text{max}} × (LW^2 B_a + M|\theta|))$.

\section{Performance Evaluation}\label{S7: Perfo Eval}

\begin{figure*}[tb]
	\centering
	\subfigtopskip=2pt 
	\subfigbottomskip=2pt 
	\subfigcapskip=-5pt 

	\subfigure[$F_n^{\text{ec}}$ v.s. $D_n^{\text{th}}$ for CF, SINR = 20dB]{
	    \label{Fig:Result0_1}
	    \includegraphics[width=5.5cm]{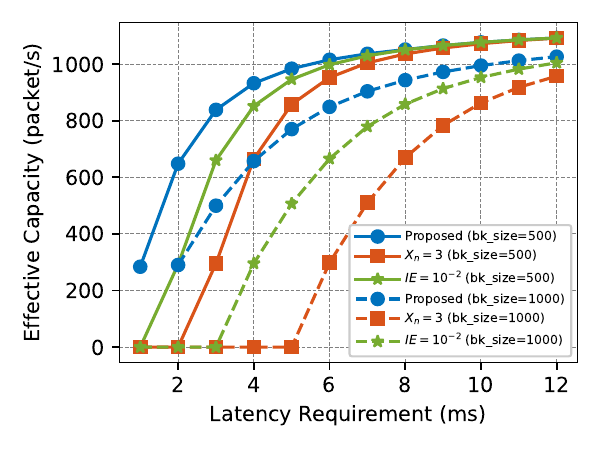}}
	\subfigure[$F_n^{\text{ec}}$ v.s. $\varepsilon_n$ for EF, SINR = 20dB]{
	    \label{Fig:Result0_2}
	    \includegraphics[width=5.5cm]{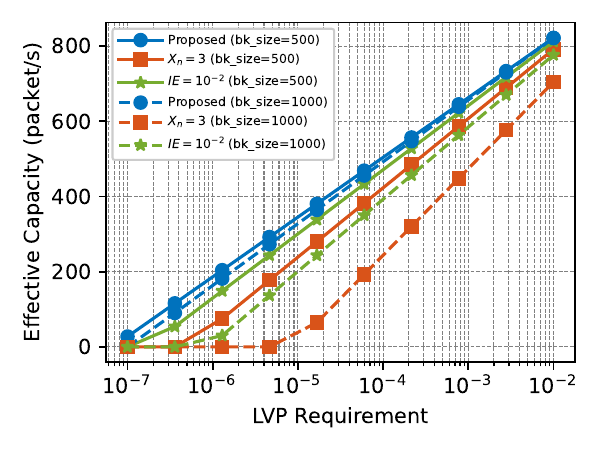}}
	\subfigure[$F_n^{\text{ec}}$ v.s. SINR, block size = 1000]{
	    \label{Fig:Result0_3}
	    \includegraphics[width=5.5cm]{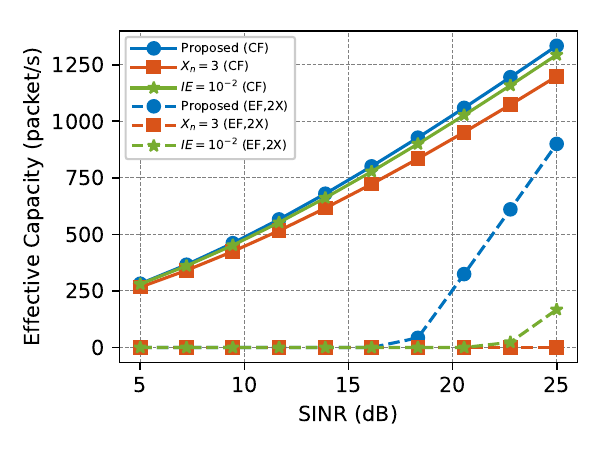}}
	\caption{The impact of coupling factors on effective capacity, where the bandwidth is 1.6 MHz, the $\varepsilon_{0} = 10^{-4}$, and the $L_p$ is 1K bytes.}
	\label{Fig:Result_0}
\end{figure*}

\subsection{Model Validation}
To characterize the coupling of transmission parameters and their combined effect on performance, and the effectiveness of our QoS analysis framework in capturing coupling relationships, a large number of controlled experiments were implemented.  Fig. \ref{Fig:Result0_1} shows the effect of total delay requirements on the effective capacity of CF at different block sizes. As the delay requirement becomes looser, the effective capacity of the system, i.e., the traffic rate with statistical QoS guarantee, quickly increases and tends to stabilize (the transmission rate of the system). Additionally, systems with longer block lengths achieve higher capacity performance, as longer block lengths can promote an increase in transmission rate or a decrease in block error rate. LVP has a similar effect on effective capacity as delay requirement, as shown in Fig. \ref{Fig:Result0_2}. Fig. \ref{Fig:Result0_3} shows the system performance resulting from different TPD schemes under different channel conditions, where the value of the effective capacity of EF is multiplied by 2 to be plotted. We can conclude that the effective capacity of CF is higher than that of EF in the same system, due to the deterioration of queuing performance by fluctuating flow in EF. The performance data in all three figures verifies the optimality of the proposed TPD scheme, which is consistent with our theoretical analysis. In addition, when the QoS requirement is too strict or the transmission conditions are poor, an effective capacity performance of 0 is obtained in three figures, which indicates that the current system cannot realize the QoS provision of this service.

\subsection{Simulation Settings of the DRL}
In our simulations, performance was evaluated on a PC equipped with an Intel(R) Core(TM) i7-10700 CPU @ 2.9 GHz and 32 GB of RAM. The software environment was based on the Anaconda platform, integrating Python 3.10 and PyTorch 2.2 within a Windows 10 Ultimate 64-bit system.

Assume 6 APs are uniformly installed within a 400 $\times$ 300 $m^2$ area, and users are randomly distributed within the coverage area of the base stations. The QoS metric tuples for users with different service demands are randomly selected within proprietary ranges. The QoS tuple range for CF users is defined as \{[$20Mbps$, $40Mbps$], [$8ms$, $15ms$], [$10^{-5}$, $10^{-3}$]\}, while the QoS range for EF is \{[$1Mbps$, $5Mbps$], [$1ms$, $5ms$], [$10^{-7}$, $10^{-5}$]\}. According to the relevant 3GPP standards, specific parameters are shown in Tab. \ref{tabPara2}. In this section, we simulate the proposed COHA with ES and compare it with different DRL schemes described below:

\begin{table}[ht]
\centering
\caption{Simulation Parameters}
\label{tabPara2}
\begin{tabular}{ll}
\toprule
Parameter                     & Value                             \\ 
\midrule
AP/UE Distribution            & Uniformly / Random                \\
Noise Power                   & \SI{-114}{dBm}                    \\
AP Max Power                  & \SI{24}{dBm}                      \\
Ant. Gain (AP/UE)             & \SI{8}{dB} / \SI{3}{dB}           \\
Bandwidth per SC              & \SI{1.6}{MHz}                     \\
Length of Slot (EF / CF)      & 0.5ms / 1ms                       \\
Packet Size (EF / CF)         & 1KB / 4KB                         \\
Path Loss Model               & $128.1 + 37.6\log_{10}(d(\mathrm{km}))$ dB \\
P.A. Efficiency               & 0.5                               \\
AP Static Power               & Half of Maximum                   \\
Decoding BLEP threshold      & $10^{-4}$                         \\
Max Transmission Count        & 5                                 \\
\bottomrule
\end{tabular}
\end{table}

\begin{itemize}
\item \textbf{COHA-based scheme: }The performance of COHA without experience sharing is meticulously measured.

\item \textbf{IPPO-based scheme: }The simplest strategy in reinforcement learning schemes involving multiple agents is to employ multiple independent decision-making and updating processes.

\item \textbf{Random-based scheme: }As a baseline comparison, we also measured the performance of random SC orchestration and power allocation to validate the effectiveness of the proposed scheme in intelligent decision-making.
\end{itemize}

In the simulation results below, we use COHA-ES and COHA to represent the scheme with experience sharing and pure COHA, respectively. And the performance of the following solutions is compared to verify the role of each part of our framework:

\begin{figure*}[tb]
	\centering
	\subfigtopskip=2pt 
	\subfigbottomskip=2pt 
	\subfigcapskip=-5pt 

	\subfigure[$\vartheta=1$: ESE reward]{
	    \label{Fig:Result1_1_band}
	    \includegraphics[width=5.5cm]{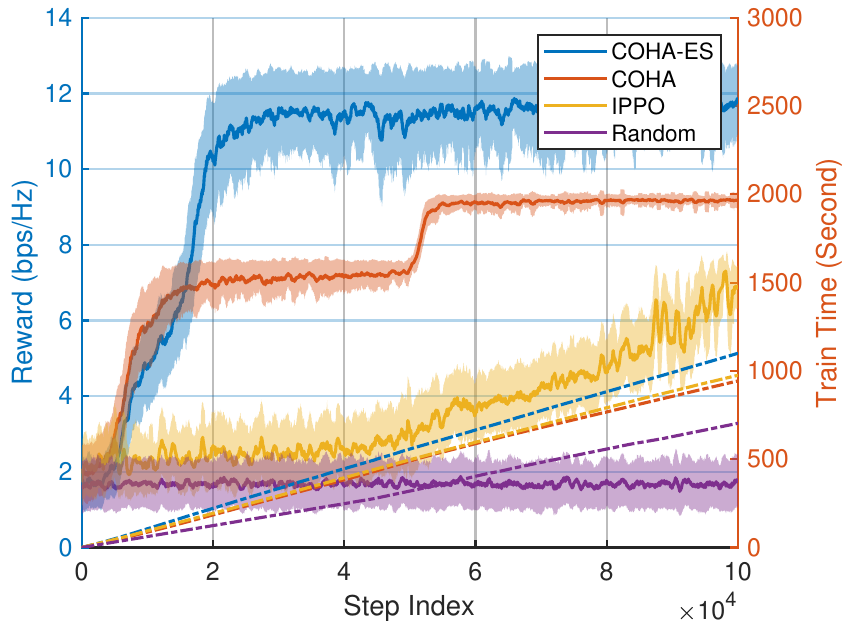}}
	\subfigure[$\vartheta=0$: EEE reward]{
	    \label{Fig:Result1_2_power}
	    \includegraphics[width=5.5cm]{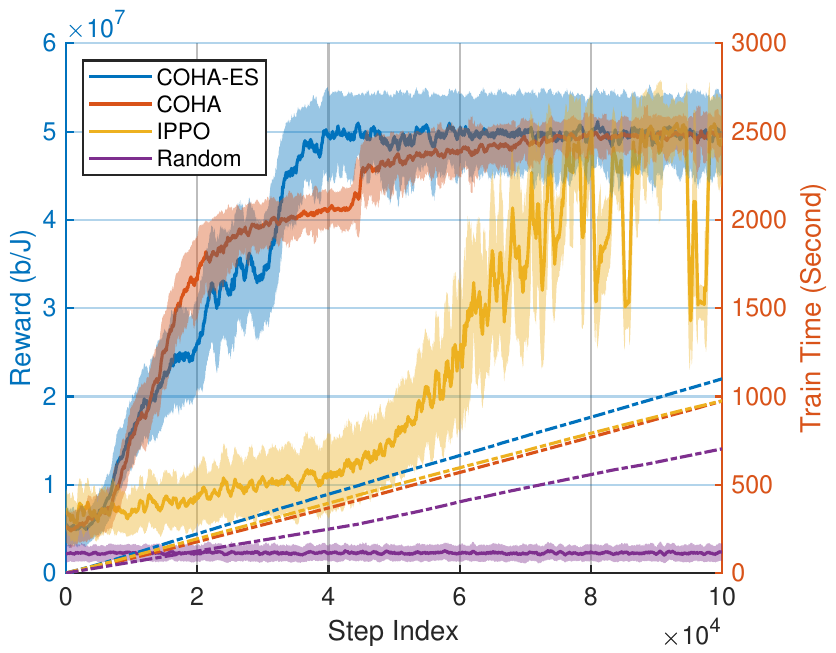}}
	\subfigure[$\vartheta=0.5$: Normalized reward]{
	    \label{Fig:Result1_3_mixed}
	    \includegraphics[width=5.5cm]{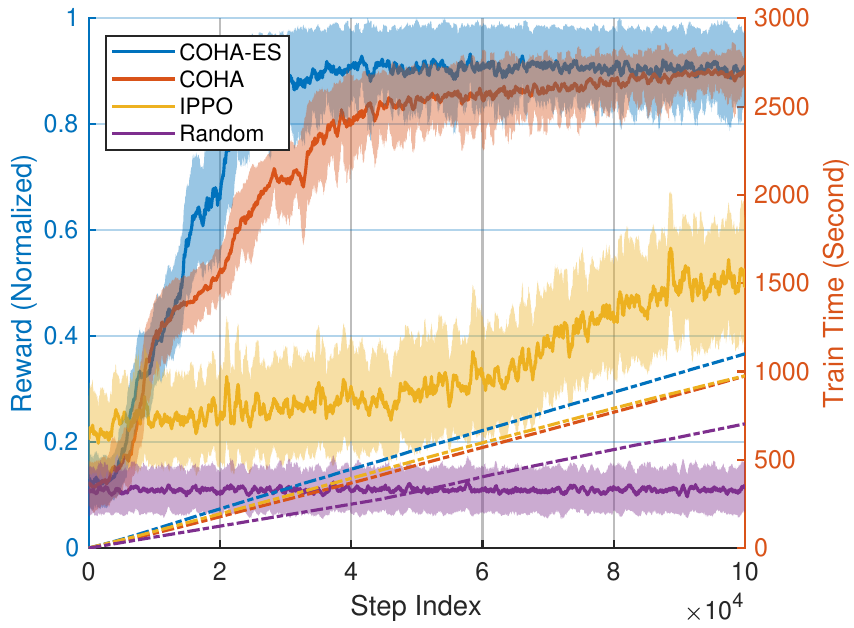}}

	\caption{Convergence performance and offline training time comparison with 5 SCs and 3 UEs.}
	\label{Fig:Result_1}
\end{figure*}

\begin{itemize}
\item \textbf{Fixed SC layout based scheme: }An SC orchestration scheme is randomly generated at the beginning of training and maintained throughout the training process, while power allocation decisions were optimized consistently with other schemes. 

\item \textbf{Average power allocation based scheme: }The maximum power of each AP is always equally distributed among the SCs, while the SC allocation and on/off decisions remain unchanged.

\item \textbf{Fixed transmission parameters based scheme: }Similar to existing communication solutions, the transmission count of data block or the decoding error probability is fixed when calculating communication capacity and reliability.

\end{itemize}

\begin{figure*}[tb]
	\centering
	\subfigtopskip=2pt 
	\subfigbottomskip=2pt 
	\subfigcapskip=-5pt 

	\subfigure[$\vartheta=1$: ESE reward]{
	    \label{Fig:Result2_1_band}
	    \includegraphics[width=5.5cm]{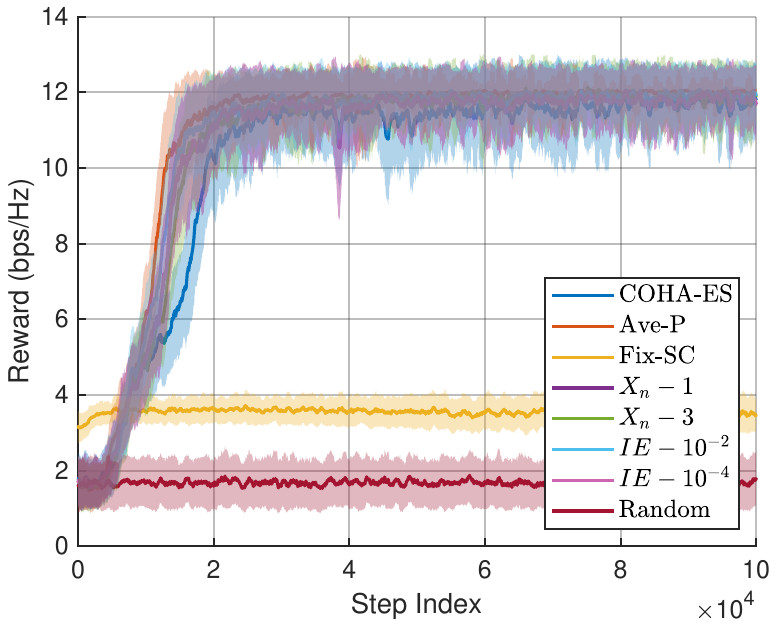}}
	\subfigure[$\vartheta=0$: EEE reward]{
	    \label{Fig:Result2_2_power}
	    \includegraphics[width=5.5cm]{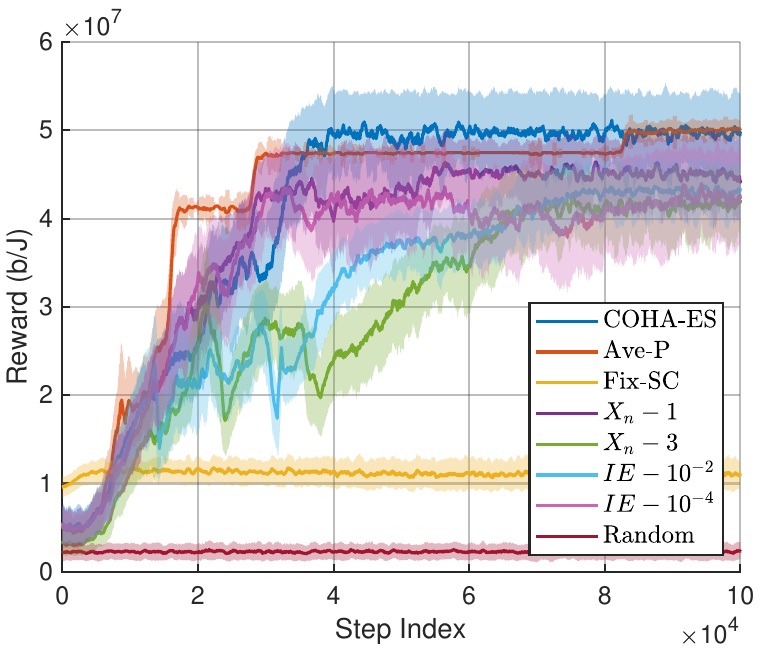}}
	\subfigure[$\vartheta=0.5$: Normalized reward]{
	    \label{Fig:Result2_3_mixed}
	    \includegraphics[width=5.5cm]{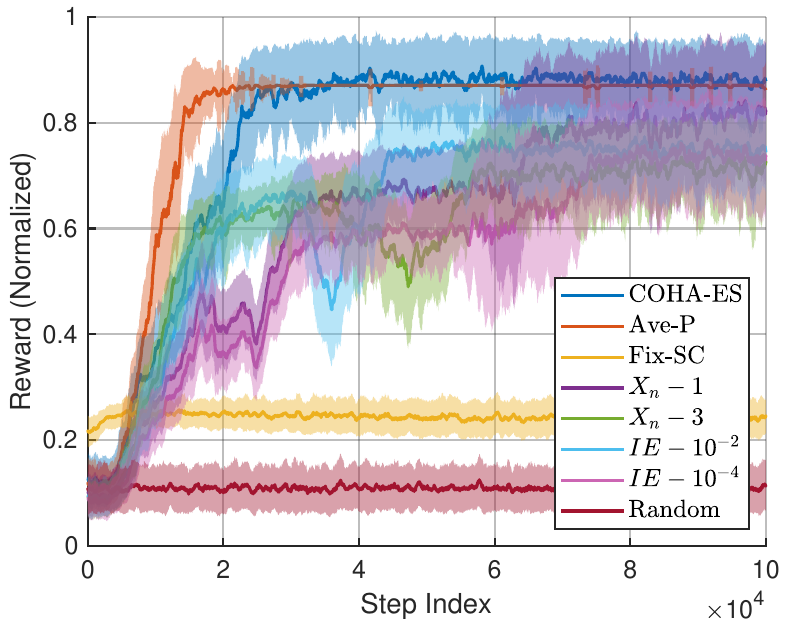}}

	\caption{Ablation experiments on individual components of the solution framework to compare convergence performance and system rewards. $Ave-P$ denotes the average power allocated to each SC; $Fix-SC$ represents a randomly determined SC orchestration at the start of training; $X_n-x$ indicates a fixed number of $x$ transmissions, while $IE-e$ denotes a fixed BLEP of $e$.}
	\label{Fig:Result_2}
\end{figure*}

\subsection{Training convergence result analysis}
Fig. \ref{Fig:Result_1} illustrates the convergence of different DRL algorithms on average reward under 5 SCs and 3 UEs, with varying emphasis on objectives. It is noteworthy that our work employs step-level reward trajectories to diagnose fine-grained policy refinement patterns, aligning with the need to characterize transient resource allocation dynamics inherent to BCN reconfiguration tasks. Alternative temporal granularities (e.g., episode-averaged metrics \cite{mohajer2024dynamic}) may better suit scenarios requiring stability-focused algorithm comparisons. As shown in Fig. \ref{Fig:Result1_1_band}, with a weighting factor $\vartheta = 1$, the system solely pursues the highest effective spectral efficiency. Conversely, when $\vartheta = 0$, effective power efficiency dominates the system reward (Fig. \ref{Fig:Result1_2_power}). Fig. \ref{Fig:Result1_3_mixed} presents the training reward for $\vartheta = 0.5$. It is noteworthy that the system rewards emphasizing a single metric (\ref{Fig:Result1_1_band}and \ref{Fig:Result1_2_power}) are not normalized to clearly characterize performance, whereas the weighted reward had to be normalized for computation. An intuitive conclusion is that system reward in pursuit of integrated performance cannot reach 1 due to the trade-off between spectral efficiency and power performance.

The results indicate that our COHA scheme significantly outperforms the IPPO method, with convergence beginning early in the training phase. The proposed experience sharing mechanism achieves substantial convergence target gains at a minimal training time cost, specifically manifested as higher convergence rewards when emphasizing only ESE (\ref{Fig:Result1_1_band}) and faster convergence speeds when emphasizing EEE (\ref{Fig:Result1_2_power} and \ref{Fig:Result1_3_mixed}). The analysis for Fig. \ref{Fig:Result1_1_band} shows that COHA scheme improves from 5 SC consumption to 4 SC consumption in the middle of training, while COHA-ES converges to the performance of 3 SC consumption in the early stage. The mutual experience sharing enables the acquisition of higher-reward training trajectories more frequently for updates. For COHA without ES, we observed significant reward fluctuations in the middle to later stages of the training process, caused by the difficulty in detecting optimal SC and AP usage.

Different from COHA and COHA-ES, IPPO fails to find an acceptable solution by the end of the training. On one hand, IPPO demonstrated the poorest stability during training, even in the later stages, as reflected by the large error bands and sudden drops shown in Fig. \ref{Fig:Result1_2_power}. On the other hand, the training process of the IPPO scheme always converged to a lower level only after prolonged periods, as indicated by the three performance-focused configurations in Fig. \ref{Fig:Result_1}. These observations indicate that agents in IPPO neither understand other decision behaviors and their impact on overall performance nor work cooperatively. This results in a failure to achieve efficient SC orchestration and power allocation, with higher time costs due to independent update process.

\subsection{Ablation analysis of frame component}
The offline training performance of ablation experiment with off-components in the system framework is shown in Fig. \ref{Fig:Result_2}. It can be observed that the COHA-ES scheme surpasses all other schemes when aiming for pure EEE performance as well as comprehensive performance. Additionally, when targeting pure ESE performance, it achieves convergence performance and system rewards consistent with all schemes except the random and the fixed SC based scheme, which is impressive and instructive. Since the bandwidth is allocated and utilized in the format of SCs, the performance loss due to component ablation in the scheme is compensated by the redundant bandwidth under 3 SCs shown in Fig. \ref{Fig:Result2_1_band}. The highly varied results in Fig. \ref{Fig:Result2_2_power} and \ref{Fig:Result2_3_mixed} confirm our analysis, as the power leading to EEE performance is allocated and consumed in an ungraded manner. Therefore, finer granularity in SC bandwidth configuration would result in better ESE performance, but it would also expand the action space of the SC actor, increasing the training burden.

Another noteworthy phenomenon is the poor performance of the scheme based on fixed SC allocation and the pretty performance of the scheme based on equal power distribution. The analysis concludes that appropriate SC usage plays a decisive role in system performance. The scheme can still flexibly control interference between APs based on suitable SC orchestration and on-off decisions under equal power allocation, while power adjustment alone is challenging to overcome the performance disadvantages under unreasonable SC allocation. Furthermore, the scheme based on equal power distribution exhibits a step-like convergence at a slower rate than the SC-power co-optimization (COHA-ES) scheme in Fig. \ref{Fig:Result2_2_power}, while it converges at the fastest rate to the optimal level with minimum variance in Fig. \ref{Fig:Result2_1_band} and \ref{Fig:Result2_3_mixed}. The faster convergence is attributed to the acceleration of the training process solely by SC actor optimization, whereas the absence of power actor and the equal power allocation adversely affect EEE performance, leading to slower convergence in Fig. \ref{Fig:Result2_2_power}.

Finally, the scheme based on fixed transmission parameters experiences various degrees of performance degradation in Fig. \ref{Fig:Result2_2_power} and \ref{Fig:Result2_3_mixed}. This is due to the inflexible fixed packet transmission count and BLEP requirements, which cannot adapt to the upper-layer services with varying reliability demands and the lower-layer transmission environments caused by changing resource allocation. The performance degradation not only results in the downgrading of the final reward but also affects the training process, as evidenced by the highly fluctuating training rewards and later convergence times in the latter two figures. As previously mentioned, the impact on ESE performance is mitigated by the bandwidth redundancy under SC configuration, which does not affect the system rewards aimed purely at ESE performance.

\begin{figure*}[tb]
	\centering
	\subfigtopskip=2pt 
	\subfigbottomskip=2pt 
	\subfigcapskip=-5pt 

	\subfigure[$\vartheta=1$: ESE reward]{
	    \label{Fig:Result3_1_band}
	    \includegraphics[width=5.5cm]{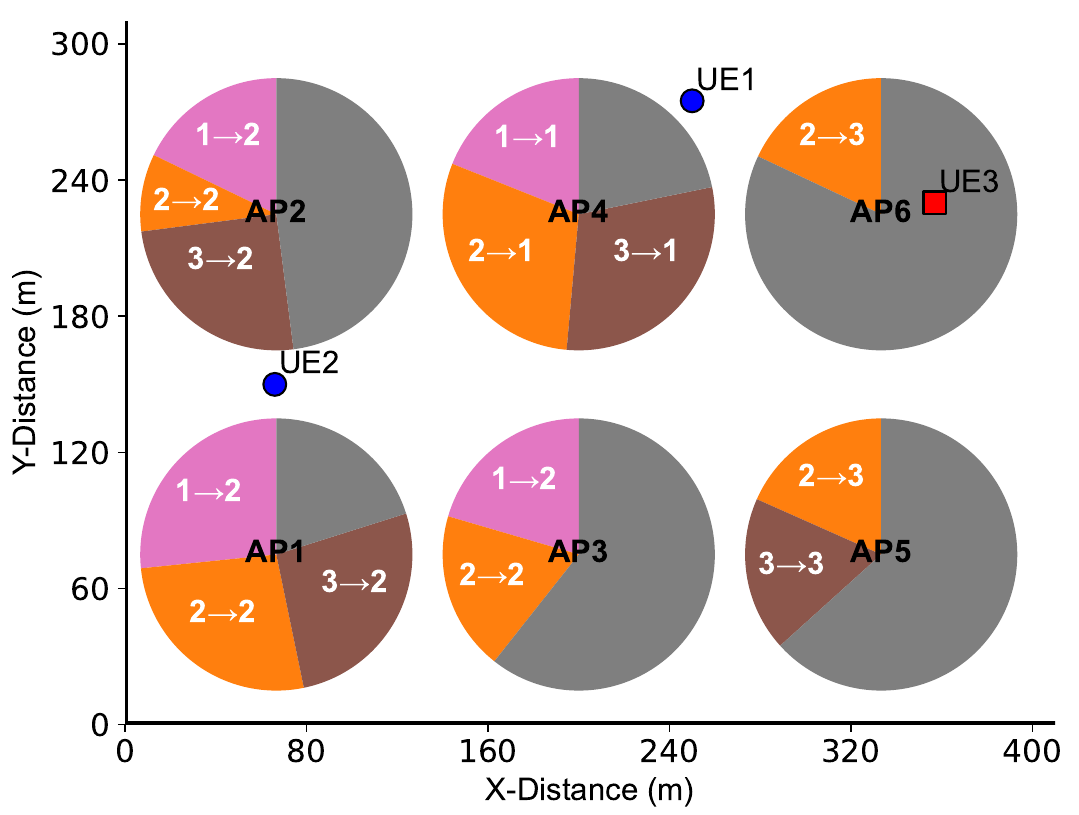}}
	\subfigure[$\vartheta=0$: EEE reward]{
	    \label{Fig:Result3_2_power}
	    \includegraphics[width=5.5cm]{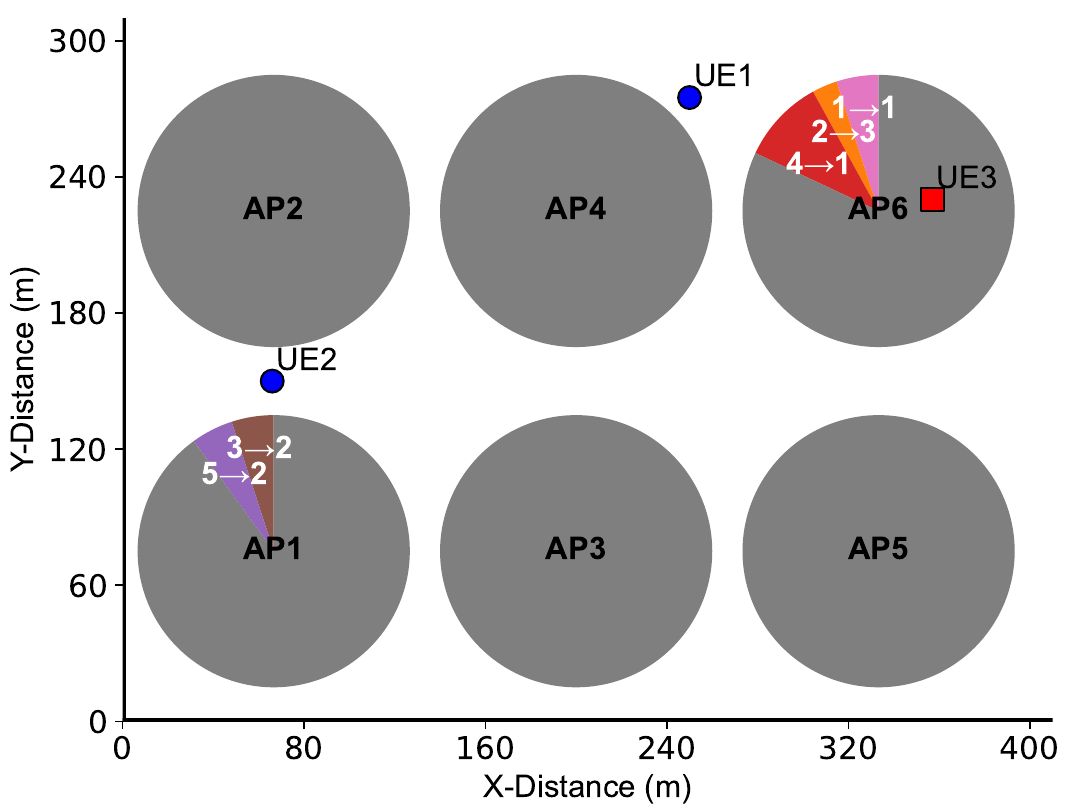}}
	\subfigure[$\vartheta=0.5$: Normalized reward]{
	    \label{Fig:Result3_3_mixed}
	    \includegraphics[width=5.5cm]{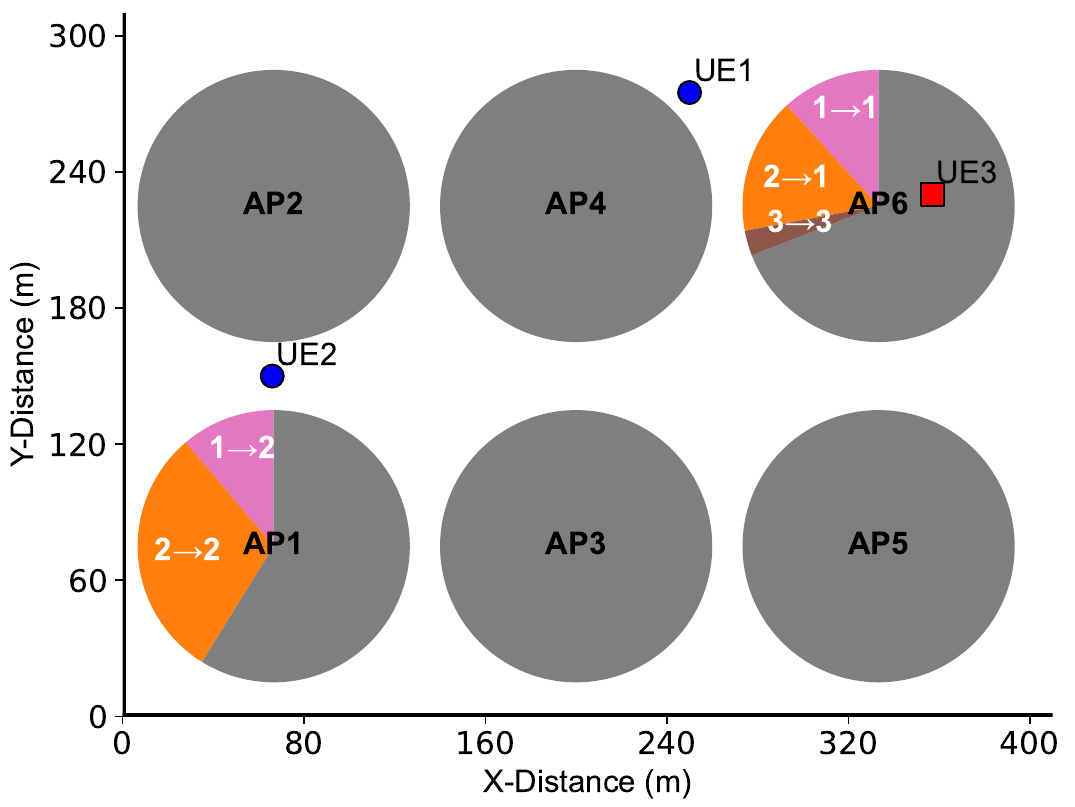}}

	\caption{SC Scheduling and power allocation decisions for each AP under different weighting factors. Circles represent CF UEs, while square users request EF UEs. Each colored sector represents a SC, with the angle indicating the power allocation on that SC. The expression $a \rightarrow b$ on a sector denotes that SC $a$ serves UE $b$. Grey sectors indicate SCs or APs with zero power allocation (i.e., in sleep mode).}
	\label{Fig:Result_3}
\end{figure*}

\subsection{Decision result analysis}
The SC scheduling and power allocation decisions made by a well-trained agent according to our performance requirements are illustrated in Fig. \ref{Fig:Result_3}. Note that the sectors do not represent the coverage area of the AP signal but are used to explain the agent's decision actions only. An intuitive result from the figure is that the EEE of the decision action when $\vartheta = 1$ is 1.67 times that when $\vartheta = 0$. Conversely, decisions solely pursuing power efficiency achieve about 4.5-fold EEE performance gain compared to decisions solely pursuing spectral efficiency with maximum SC consumption. When compared to decisions aiming for integrated performance ($\vartheta = 0.5$), this gain is 150\%. When solely pursuing spectral efficiency ($\vartheta = 1$), the system minimizes SC usage to enhance spectral efficiency, which requires activating multiple APs to improve channel quality at the expense of reduced energy efficiency. As observable in Fig. \ref{Fig:Result3_1_band}, APs reusing SCs consistently prioritize serving nearby users. This strategy simultaneously enhances the channel quality of served users while reducing interference to others, effectively demonstrating both the validity of our interference model and its impact on resource orchestration.

The decision results effectively demonstrate that a properly designed agent can deeply understand the service requirements and make sound decisions based on performance objectives. The APs consistently tend to serve users who are in close proximity, and the utilization of the same SC by multiple APs is well-coordinated to achieve a satisfactory balance between spectral efficiency and interference. An intriguing result is that decisions made under comprehensive performance objectives, as opposed to configurations solely pursuing spectral efficiency, do not increase the usage of SCs while maintaining efficacy, which can be explained from two perspectives. First, the redundant bandwidth under the SC mechanism compensates for the increased bandwidth demand due to efficiency requirements, as previously mentioned. Second, the appropriate SC scheduling itself is conducive to improving power efficiency, which will be further analyzed in subsequent sections. From the perspective of application types, EF services (UE $3$) are configured with stricter reliability requirements, which drive reduced SC reuse, as illustrated in Fig. \ref{Fig:Result3_3_mixed} where SC $3$ is exclusively utilized by UE $3$. In contrast, this phenomenon is absent in Fig. \ref{Fig:Result3_1_band} and \ref{Fig:Result3_2_power} due to the urgent pursuit or complete relaxation of spectral efficiency. This result demonstrates the capability of model to capture service reliability requirements.

\begin{figure}
    \centering
    \subfigtopskip=2pt 
    \includegraphics[width=7cm]{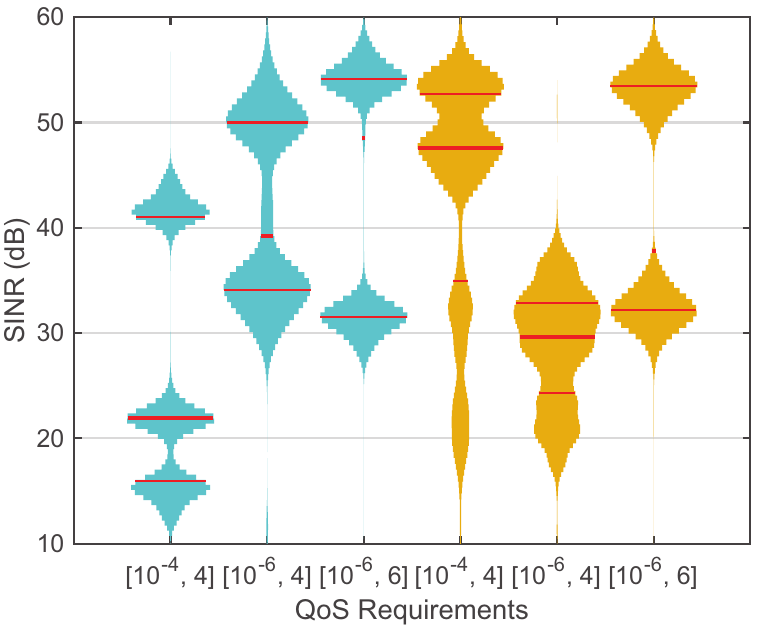}
    \caption{SINR distribution of users on SCs when serving CF (cyan) and EFs (yellow) with different QoS requirements, where the labels $[x, y]$ on the X-axis represent the LVP and delay (ms) restrictions of the flow respectively. The red lines in the plot represent the 25\%, 50\% and 75\% quantiles of the distribution.}
    \label{Fig: Result4}
    \vspace{-0.35cm} 
\end{figure}

To validate the capability of the resource allocation scheme in perceiving application types and QoS requirements, we measured the channel quality distribution (SINR) of users with application flows subjected to different QoS constraints, as illustrated in the violin plots in Fig. \ref{Fig: Result4}. The results were obtained in a signal environment influenced by the decisions made by agents pursuing EEE. The most direct observation is the distinct shapes and positions of the violin plots for CFs and EFs, indicating significant differences in the SINR distributions of users under different QoS requirements. A prominent area in each violin plot represents a SINR concentration, which is a result of the combined effects of SC orchestration and power allocation leading to superposition of useful and interfering signals. Further analysis reveals that the SINR distribution for EFs is more dispersed compared to CFs, suggesting that EFs can achieve high system performance across a wider range of channel environments through the regulation of transmission parameters. In summary, the proposed framework can accurately capture the QoS requirements and locations of users and make optimal resource allocation decisions based on this information.

\subsection{Effect of scenario parameters on performance}
\begin{figure}
    \centering
    \includegraphics[width=7cm]{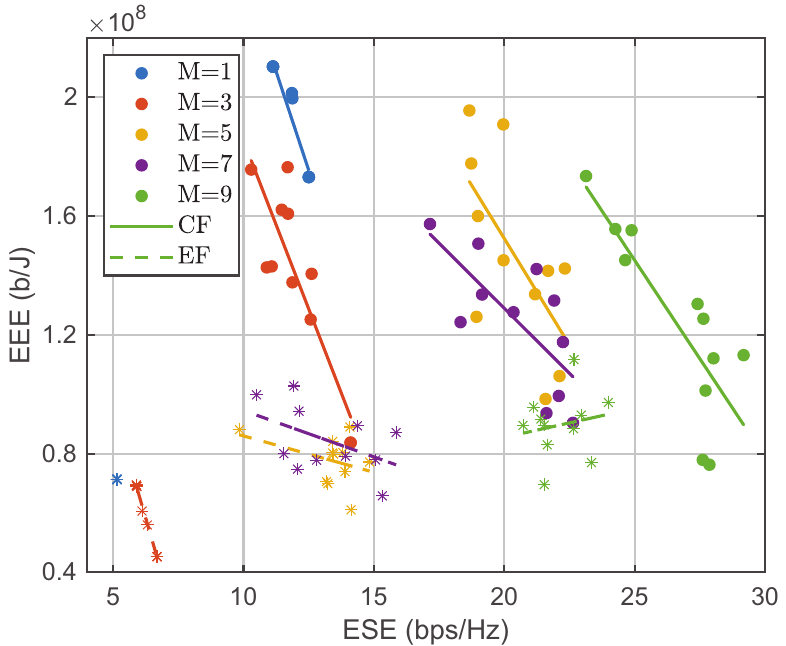}
    \caption{EEE-ESE trade-off curve with varying number of APs and fixed 3 UEs and 5 SCs.}
    \label{Fig: Result5_1}
\end{figure}

\begin{figure}
    \centering
    \includegraphics[width=7cm]{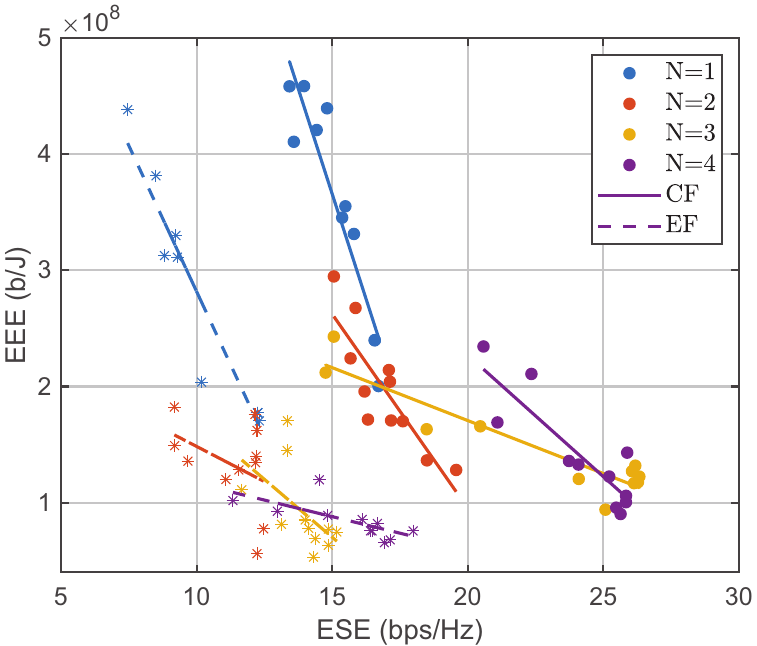}
    \caption{EEE-ESE trade-off curve with varying number of UEs and fixed 6 APs and 5 SCs.}
    \label{Fig: Result5_2}
\end{figure}

A series of extensive random trials were conducted to evaluate the performance of the BCN architecture. The outcomes are illustrated in Fig. \ref{Fig: Result5_1} and \ref{Fig: Result5_2}. Typically, more stringent latency and reliability constraints are imposed on EFs, consistent with the inherent characteristics of both CFs and EFs. By applying distinct weighting factors $\vartheta$ to the EEE and ESE scores, the performance trade-off of EEE-ESE under a fixed configuration was examined. Notably, divergent conclusions regarding CFs and EFs were drawn from the varying AP results, as evidenced in Fig. \ref{Fig: Result5_1}. Similar conclusions were observed in the results concerning the varying number of UEs, as depicted in Fig. \ref{Fig: Result5_2}.

The observation from Fig. \ref{Fig: Result5_1} reveals that the system's ESE performance improves rapidly with the increase in the number of APs. Specifically, the ESE performance for CFs sees a maximum enhancement of 162\% with a nine-AP configuration compared to a single-AP setup, while EFs achieve a maximum improvement of 366\%. The enhancement in ESE performance for CFs hurts EEE performance, which is not observed for EFs. For each AP configuration, a mutually exclusive relationship is observed between the EEE and ESE performance of CFs. This is characterized by the trade-off, where an improvement in one performance metric invariably results in a decrease in the other. However, the conclusions differ for EFs. As the number of APs in the system increases, the EEE and ESE performance of EFs under the same AP configuration transition from a competitive relationship to a synergistic improvement. The fundamental cause of these performance differences lies in the increased capacity and the additional power consumption incurred by activating more APs. With the increase in the number of APs, the additional power consumption becomes non-negligible as more APs are activated. While the total system capacity increases, so does the level of interference. The critical factor is that the capacity of EFs improves more rapidly compared to CFs due to the relationship between its effective capacity and channel capacity. The rapid enhancement of the effective capacity for EFs lead to a significant improvement in overall performance, driven by increased emphasis on EEE capabilities and the proliferation of APs.

We then discuss the impact of varying the number of users in the scenario on system performance. Notably, under a configuration with six APs, the EEE performance of EFs remains competitive with ESE performance, resulting in a consistent performance trend similar to that of CFs, as illustrated in Fig. \ref{Fig: Result5_2}. In the experiments conducted, the ESE performance of CFs increased by up to 96.06\% with an increase in the number of users, while EFs saw a maximum increase of 141.05\% , albeit with a corresponding 5.1x and 8.2x degradation in EEE performance, respectively. The implementation of more advanced interference avoidance techniques is expected to shift the system performance curve further towards the upper right, indicating an overall improvement in EEE-ESE performance. In summary, the performance trade-off curves demonstrate the exceptional capability of our BCN framework in adapting transmission schemes based on performance objectives.

\section{Conclusion}\label{S8: Conclu}
Unlike existing communication systems that offer standardized QoS provision for different application scenarios, we investigate a business-centric resource orchestration and transmission parameter decision framework. The proposed scheme adaptively constructs optimal channels and determines transmission parameters for various types of applications with different QoS requirements, thereby maximizing weighted resource efficiency while achieving statistical QoS provision. To address the challenges of large action spaces and coupled variables, we design a novel collaborative DRL algorithm with experience sharing. Extensive experimental results demonstrate that our algorithm achieves the fastest convergence speed and obtains the highest rewards compared to other algorithms.

The statistical QoS guarantee model in our study characterizes the service capacity that can be provided with a certain QoS guarantee in a fluctuating channel environment, which implies a relatively stable source rate or its distribution. Providing a QoS guarantee for highly fluctuating traffic load is the limitation of our model, which can serve as future research. In addition, the data processing capacity of each AP is actually limited, and the AP model considering certain capacity or different power constraints can be studied in future work.

\bibliographystyle{IEEEtran}

\bibliography{202310}

\end{document}